\documentclass[letterpaper,12pt]{article}
\usepackage{amsmath}
\usepackage{amssymb}
\usepackage{mathtools}
\usepackage{graphicx}
\usepackage{algorithm}
\usepackage{algpseudocode}
\usepackage{algorithmicx}
\usepackage{cprotect}
\usepackage{multirow}
\usepackage{jheppub}

\usepackage{adjustbox}
\usepackage{tikz}
\usetikzlibrary{quantikz}
\usetikzlibrary{external}

\newcommand{\Id}{\mathit{Id}}

\NewDocumentCommand{\bbeta}{o}{
	\pmb{\beta}\IfValueT{#1}{_{[#1]}}
}
\NewDocumentCommand{\bgamma}{o}{
	\pmb{\gamma}\IfValueT{#1}{_{[#1]}}
}
\NewDocumentCommand{\bgstate}{o}{
	\ket{	\IfValueTF{#1}{\bbeta[#1]}{\bbeta},
		\IfValueTF{#1}{\bgamma[#1]}{\bgamma}
	}
}
\NewDocumentCommand{\bgstateT}{o}{
	\bra{	\IfValueTF{#1}{\bbeta[#1]}{\bbeta},
		\IfValueTF{#1}{\bgamma[#1]}{\bgamma}
	}
}

\usepackage{hyperref}
\usepackage{doi}

\title{Fair Sampling Error Analysis on NISQ Devices}
\preprint{LA-UR-21-20101}

\author{John~Golden,$^{1,2,\dagger}$\note[$\dagger$]{Corresponding author: golden@lanl.gov}}
\author{Andreas~Bärtschi,$^1$}
\author{Daniel~O'Malley,$^2$}
\author{Stephan~Eidenbenz$^1$}

\affiliation{$^1$ Information Sciences (CCS-3), Los Alamos National Laboratory, Los Alamos, NM 87545}

\affiliation{$^2$ Computational Earth Sciences (EES-16), Los Alamos National Laboratory, Los Alamos, NM 87545}

\begin{document}

\abstract{
We study the status of \emph{fair sampling} on Noisy Intermediate Scale Quantum (NISQ) devices, in particular the IBM Q family of backends. 
Using the recently introduced Grover Mixer-QAOA algorithm for discrete optimization, we generate fair sampling circuits to solve six problems of varying difficulty, each with several optimal solutions, which we then run on twenty backends across the IBM Q system.
For a given circuit evaluated on a specific set of qubits, we evaluate: how frequently the qubits return an optimal solution to the problem, the fairness with which the qubits sample from all optimal solutions, and the reported hardware error rate of the qubits. 
To quantify fairness, we define a novel metric based on Pearson's $\chi^2$ test. 
We find that fairness is relatively high for circuits with small and large error rates, but drops for circuits with medium error rates. 
This indicates that structured errors dominate in this regime, while unstructured errors, which are random and thus inherently fair, dominate in noisier qubits and longer circuits.
Our results show that fairness can be a powerful tool for understanding the intricate web of errors affecting current NISQ hardware. 
}

\maketitle
\section{Introduction}
Combinatorial optimization problems, such as the Minimum Traveling Salesperson problem, the Maximum Satisfiability problem, or the Spin Glass Ising model, are widely viewed to be one of the most promising application domains for quantum computing. 
Such problems often have multiple optimal solutions, and depending on the algorithm being used to solve the problem, some optima may be much easier to find than others. 
This is problematic if one is interested not in finding a single solution, but instead determining the overall landscape of optimal solutions.
Algorithms which fairly sample from the space of optimal solutions avoid this problem and have practical value in physics and engineering.
For instance, it is often difficult in practice to encode all the engineering design goals in the function to be optimized.
By allowing an engineer to explore an unbiased assortment of optimal solutions, fair sampling provides an opportunity to apply these other design goals to a more diverse set of options (especially useful when a customer adds one more requirement after the optimization has been performed).
It can also be used to produce an ensemble of predictions from calibrated physical models that can be used to make a forecast including a range of possible outcomes (e.g., in the context of fluid flow in the earth's subsurface, which can be formulated as a discrete optimization problem \cite{harp2008aquifer,o2018approach}).

In this paper, we determine the current status of fair sampling in the context of quantum computing.
Specifically, we are interested both in algorithms which theoretically guarantee fair sampling (i.e. generate quantum states where all optimal solutions have the same amplitudes squared and thus are measured with equal probability), as well as actual hardware implementations of such algorithms. 
Previous studies have shown that quantum annealing cannot be used as a fair sampler for all problems~\cite{K_nz_2019}. 
We therefore focus our work on gate-based quantum computing algorithms and hardware.

The most widely-used quantum optimization framework is the Quantum Alternating Operator Ansatz (QAOA), which alternates between a mixer operator that mixes the amplitudes of all feasible solutions and a phase separator operator that separates the phases of the feasible solutions~\cite{hadfield_qaoa}. 
Most mixers do not exhibit fair sampling, however, the recently discovered Grover Mixer~\cite{baertschi2020grover} has the property that all feasible solutions with the same objective value have identical amplitudes. 
This novel Grover Mixer implementation of QAOA thus solves the theoretical side of fair sampling for gate-based quantum computing.

The majority of this work is dedicated to a practical evaluation of how fair sampling circuits are affected by the errors inherent in Noisy Intermediate Scale Quantum (NISQ) devices.
In this study, a paradox quickly becomes apparent: while an error-corrected quantum device could obviously be used as a perfectly fair sampler, so could a quantum device completely dominated by random noise.
Of course the noisy machine would be no different than a random number generator, and would sample from all feasible solutions rather than just the optimal solutions, but it would do so fairly.
Therefore, our study seeks to clarify where current NISQ devices lie on the spectrum between these two extremes.
Specifically, are the hardware errors predominantly random and unstructured, thus leading to relatively fair sampling of optimal solutions?
Or do hardware biases and other structured errors play an important role, thus leading to unfair behavior?

Our methodology for answering these questions consists of the following steps:
\begin{enumerate}
\item
We define six example problems, which are a slightly expanded set of examples studied in the context of fair sampling in quantum annealing~\cite{K_nz_2019}. 

\item
We design quantum circuits for our problems and hand-tune them to minimize circuit depth and gate counts, taking into account different qubit connectivity topologies as encountered on IBM Q. 
Our circuits involve two to five qubits and our medium size examples result in circuits with gate counts in the $30-60$ range, which is similar to the size of the (random) circuits used to establish the Quantum Volume metric on the IBM Q backends 

\item
We execute our circuits in a large number of shots on twenty different IBM Q backends that have qubit counts of up to 65 qubits. 
In addition, we embed our circuits in all possible topologies on the circuits resulting a total of 7440 experiments at 40960 shots each.

\item We define a novel metric for fairness in quantum computing based on Pearson's $\chi^2$ test, a common statistical tool to evaluate how likely it is that a set of observational data came from a null hypothesis distribution.
Since all NISQ hardware will necessarily involve some biases and imperfections that inhibit fair sampling, we are interested in how much effort must be expended to prove that a piece of hardware is biased, i.e. reject the null hypothesis of fair sampling. 
We term this metric `number of shots to reject fair sampling,' and it depends on both a specific set of qubits as well as a fair sampling circuit. This fairness metric is a separate contribution to the main results of the paper

\item
We analyze our experiments using our fairness metric and we calculate the aggregate error of each experiment from IBM's reported gate-specific error rates.
\end{enumerate}

We find that the NISQ devices under evaluation provide noticeably unfair sampling. 
However, fairness increases as aggregate error approaches both 0 and 1, matching with the theoretical extremes of error-corrected and noise-dominated devices discussed previously. 
In the case of small aggregate error, this is to be expected, as for 0 error we would retrieve the theoretical guarantees of Grover Mixer QAOA.
Likewise, for large aggregate error, one should get samples uniformly at random from all computational basis states.
Nonetheless, the result is perhaps counter-intuitive, as one might expect a linear interpolation between these two extremes. 
For example, the hardware-efficient random circuits in Google's recent quantum supremacy experiment~\cite{arute2019quantum} give a sample distribution consisting of a convex combination of an ideal (error-free) Porter-Thomas distribution and a Uniform distribution (caused by errors). 
Our differing results indicate that we should not expect errors themselves to be evenly distributed (i.e., in a fair manner) across qubits for highly structured circuits such as ours.
Indeed, a previous study found that structured errors (e.g. crosstalk, coherent noise) play a large role in the performance of structured circuits~\cite{proctor2020measuring}. 
Our result holds for our selected set of circuits and across the twenty NISQ backends on IBM Q.

We propose that fair sampling be an addition in the emerging set of benchmarks for NISQ devices, which so far have largely focused on circuit correctness.
As we show in this work, biased errors can artificially inflate circuit accuracy, and fair sampling on its own can be fooled by devices dominated by random noise. 
However, by studying both accuracy and fairness, we gain a more complete picture of the type and magnitude of errors affecting NISQ hardware. 

In summary, the main contributions of our work are:
\begin{enumerate}
	\item We propose a novel methodology for quantifying the fairness of sampling on NISQ hardware.
	\item We elucidate two noise regimes (high noise and low noise) that result in fair sampling, and a third regime (intermediate noise) that is dominated by biased errors.
	\item In practice, this approach provides a way to determine whether near-term improvements that reduce noise will improve or degrade the performance of fair sampling for a given problem.
	\item As an example, we study how a practical error mitigation technique (namely, measurement error mitigation) impacts the quality of sampling on IBM Q hardware.
\end{enumerate}

\section{Review}
In this section we describe previous studies of fair sampling in the context of quantum computing. 
A note on nomenclature: since the optimization problems we study can all be phrased in terms of Ising models, we adopt the more physics-oriented language and refer to `optimal solutions' as ground states. 
Problems with multiple optimal solutions are thus said to have degenerate ground states. 

\subsection{Fair sampling in quantum annealing}
After quantum annealing was proposed as a possibly useful optimization heuristic \cite{kadowaki1998quantum}, it was discovered that na\"ive approaches to quantum annealing do not always sample degenerate ground states fairly \cite{matsuda2009ground}.
In fact, certain Hamiltonians feature ground states that are sampled with probability approaching zero as annealing time increases.
This effect is inherent to the quantum annealing approaches, and is not the consequence of hardware noise or other technical limitations.
This behavior was subsequently verified experimentally \cite{mandra2017exponentially}.
Since unfair sampling (especially when certain ground states are strongly suppressed) can have negative consequences for many applications, techniques were proposed to improve upon the na\"ive quantum annealing approach \cite{sieberer2018programmable,K_nz_2019,yamamoto2020fair,kumar2020achieving}.
None of these techniques have been tested on quantum annealing hardware, indicating some of the challenges with these approaches.
It should also be noted that it has been proposed that there are some advantages of biased samplers, that is, they can be used effectively in an ensemble of samplers each of which has different biases \cite{zhang2017advantages}.

One technique that has been proposed to improve the fairness of quantum annealing is the introduction of more complex Hamiltionians to drive the annealing process \cite{mandra2017exponentially}.
This generally improves fairness, but is hard to implement and does not resolve the underlying problem.
Even for small problems, biases sometimes remain despite using fairly high-order Hamiltonians \cite{K_nz_2019}.
In studying these higher-order Hamiltonians, K\"{o}nz et al \cite{K_nz_2019} introduced four Ising models that are suitable for our QAOA analyses.
The problems are described in Table~\ref{tab:models} as Problems (a)--(d). 
These problems all exhibit biased sampling with quantum annealing, so they present a non-trivial challenge for fair sampling algorithms.
Two of them involve 4 qubits, one involves 5 qubits, and another involves 6 qubits.
This small number of qubits make it plausible that these problems can be solved with reasonably high fidelity on current NISQ hardware.

\subparagraph{Impact of noise:} Recent work~\cite{vuffray2020programmable} achieved high-quality thermal Gibbs Sampling with the D-Wave 2000Q system for a restricted class of Ising Hamiltonians that are native to the hardware connectivity and have coupling strengths $J_{ij} \in \{-1,0,1\}$. This category includes Hamiltonians with degenerate ground states, but not the more general Hamiltonians studied in this paper (see Table~\ref{tab:models}). Fair Sampling of those ground states is achieved by re-scaling the input energy scale to a sweet spot that is large enough to be noise-resilient but small enough to overcome the biases from the transverse field. In the identified sweet spot energy scale, changing the annealing time influences the effective temperature of the Gibbs samples generated by the hardware, and thus the probability to find ground states.

\subsection{Fair sampling in the Quantum Alternating Operator Ansatz}\label{sec:review-QAOA}

The Quantum Alternating Operator Ansatz~\cite{hadfield_qaoa} (QAOA) is a promising heuristic quantum algorithm for (combinatorial) optimization problems.
In its essence, for a problem instance $I$ with feasible states $F$ and cost Hamiltonian $H_C$ on $n$ qubits, a $p$-round QAOA prepares a parametrized state
from which one would like to sample low-energy states with respect to $H_C$:
\begin{align}
	\bgstate := U_M(\beta_p) U_P(\gamma_p) \cdots U_M(\beta_1) U_P(\gamma_1) U_S \ket{\uparrow^n}.	\label{eq:bgstate}
\end{align}

\begin{figure*}[t!]
	\centering
		\begin{adjustbox}{width=\linewidth}
	\newcommand{\zgate}[1]{\gate{Z^{-\beta_{#1}/\pi}}}	
	\begin{quantikz}[row sep={24pt,between origins},execute at end picture={
				\node[xshift=-10pt, yshift=-20pt] at (\tikzcdmatrixname-4-2) {
					\smash{\textcolor{red}{$\frac{1}{\sqrt{|F|}}\sum\limits_{x\in F} \ket{x}$}}
				};	
				\node[xshift=10pt, yshift=-24pt] at (\tikzcdmatrixname-4-5) {
					$\underbrace{\hspace*{310pt}}_{p\text{ rounds with angles } \gamma_1,\beta_1,\ldots,\gamma_p,\beta_p}$
				};
			}]
		\lstick{\ket{\uparrow}}	
		& \gate[4]{U_S}\slice{}
		& \gate[4]{U_P(\gamma_k)=e^{-i\gamma_k H_C}}
		& \gate[4]{U_S^{\smash{\dagger}}}
		\gategroup[4,steps=5,style={dashed,rounded corners,fill=blue!20, inner sep=0pt},background]{$U_M(\beta_k) = e^{-i\beta_k \ket{F}\bra{F}}$}
		& \targ{}
		& \ctrl{1}
		& \targ{}
		& \gate[4]{U_S}
		& \qw\midstick[4,brackets=none]{\ldots}
		& \meter{}\rstick[4]{\rotatebox{90}{$\bgstateT H_C \bgstate$}}
		\\	
		\lstick{\ket{\uparrow}}	&&&	& \targ{} 	& \ctrl{1}	& \targ{} 	&	&\qw	& \meter{}		\\	
		\lstick{\ket{\uparrow}}	&&&	& \targ{}	& \ctrl{1}	& \targ{}	&	&\qw	& \meter{}		\\	
		\lstick{\ket{\uparrow}}	&&&	& \targ{}	& \zgate{k}	& \targ{}	&	&\qw	& \meter{}	
	\end{quantikz}
	\end{adjustbox}
	\caption{Grover Mixer QAOA: The state preparation unitary $U_S$ for an equal superposition of all feasible states $\ket{F} = \frac{1}{\sqrt{|F|}}\sum_{x\in F} \ket{x}$,
		its conjugate transpose $U_S^{\dagger}$, and a multi-controlled phase-shift gate 
		$Z^{-\beta/\pi} =  \left( \protect\begin{smallmatrix} 1&0\protect\\ 0& e^{-i\beta} \protect\end{smallmatrix} \right)$ 
		are used to implement the mixer $U_M(\beta) = e^{-i\beta \ket{F}\bra{F}}$.
	}
	\label{fig:gmqaoa}
\end{figure*}
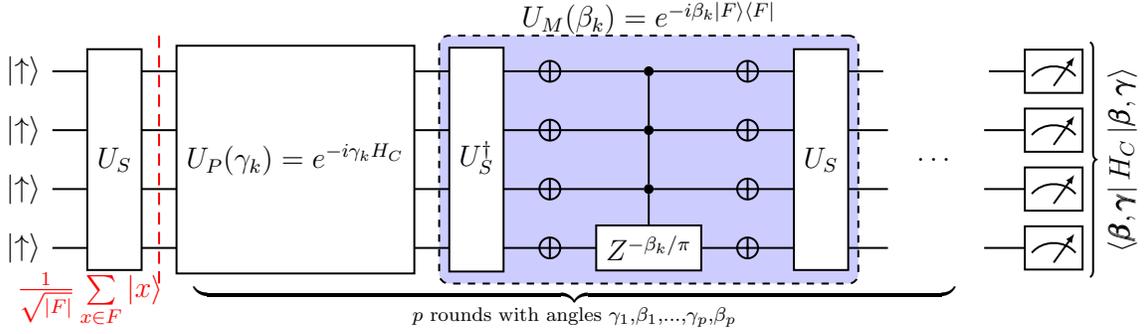

The circuit consists of an initial \emph{state preparation} unitary operator $U_S$ that creates some superposition of all feasible solutions in $I$, followed by $p$ applications of alternating parametrized \emph{phase separating} and \emph{mixing} unitaries $U_P(\gamma_k)$, $U_M(\beta_k)$ with real angle parameters $\bgamma = (\gamma_1, \ldots, \gamma_p)^T$ and $\bbeta = (\beta_1, \ldots, \beta_p)^T$, and a final measurement in the computational basis, see e.g.~Fig.~\ref{fig:gmqaoa}.
The $\bbeta, \bgamma$ angles are traditionally found via a hybrid classical/quantum approach: given some initial parameters, take enough samples to reliably estimate the expectation value $\bgstateT H_C \bgstate$, use a classical optimizer to adjust the parameters, and repeat until one has a state with a good (low-energy) expectation value. 
For further description of, and theoretical justifications for this approach, see ~\cite{Farhi2014,nasa2020XY,cook2020kVC}.

The role of the phase separating unitaries $U_P(\gamma)$ is to add multiplicative phase factors to the amplitudes of feasible computational basis states, with phases proportional to respective energies. 
We usually have (up to global phases) $U_P(\gamma) \cong e^{-i\gamma H_C}$. 
For many problems, such as MaxCut~\cite{Farhi2014} or the problems considered in this paper, $H_C$ is an Ising Hamiltonian with quadratic and sometimes linear terms; but in other problems, $H_C$ can also involve higher-order terms, such as for MaxE3Lin2~\cite{Farhi2015}.
The interplay between the state preparation unitary $U_S$ and the mixing unitary $U_M$ is particularly important, and falls into three categories:
\begin{description}
	\item[Original Approach]	
		The original approach by Farhi et.al.~\cite{Farhi2014}, termed Quantum Approximate Optimization Algorithm,
		was inspired by quantum annealing for unconstrained optimization problems, i.e. 
		$U_S = H^{\otimes n}$ (preparing an equal superposition of all states) and 
		$U_M(\beta) = e^{-i\beta \sum X_j}$ (a transverse field of Pauli-X operators acting on individual qubits).
	\item[General Ansatz] 	
		The general framework introduced in Eq.~\ref{eq:bgstate} by Hadfield et.al.~\cite{hadfield_qaoa} is a generalization
		also suitable for a wide variety of constrained optimization problems. They focus on designing involved mixing unitaries $U_M(\beta)$ and 
		usually split $U_S$ in a layer of Identity and Pauli-X gates preparing some feasible computational basis state, followed by an initial 
		mixing unitary $U_M(\beta_0)$ with new angle~$\beta_0$.
	\item[Grover Mixer]	
		If one can design efficient state preparation unitaries $U_S$ preparing an equal superposition 
		of all feasible basis states $\ket{F} = 1/\sqrt{|F|} \sum_{x\in F} \ket{x}$, then this gives rise to a mixing unitary 
		resembling Grover's selective phase-shift operator~\cite{grover2005fixed,yoder2014fixed,biamonte2018variational}, 
		$U_M(\beta) = e^{-i\beta \ket{F}\bra{F}} 
		= \Id - (1-e^{-i\beta})\ket{F}\bra{F} 
		= U_S(\Id - (1-e^{-i\beta})\ket{\uparrow}\bra{\uparrow})U_S^{\dagger}$, 
		see Fig.~\ref{fig:gmqaoa}.
		In some sense, this is the reverse of the method in the general ansatz, and it was shown to be applicable to
		many relevant optimization problems~\cite{baertschi2020grover}, 
		but there also exists a range of constrained problems where it is ruled out under complexity-theoretic assumptions~\cite{baertschi2021grover}. 
\end{description}

Although the Quantum Alternating Operator Ansatz is the most general framework of the three, both the Quantum Approximate Optimization Algorithm as well as Grover Mixer QAOA have their advantages. 
For example, the original approach by definition can not do worse than a Trotterization of the adiabatic algorithm~\cite{Farhi2000}, thus converging to a ground state in the $p \rightarrow \infty$ limit~\cite{Farhi2014}. 
Furthermore its phase separating and mixing operators are periodic, restricting the angles to real values in $[-\pi,\pi)$. 
Grover Mixer QAOA retrieves these properties for constrained optimization problems with efficient state preparation unitaries $U_S$~\cite{baertschi2021grover}, but for unconstrained optimization problems the Grover Mixer $e^{-i\beta \ket{F}\bra{F}}$ has a larger circuit depth than the Transverse Field Mixer $e^{-i\beta \sum X_j}$.
In this paper we explore an advantage unique to Grover Mixer QAOA among the three categories: fair sampling.

With Grover Mixer QAOA, all feasible basis states begin with amplitude $1/\sqrt{|F|}$ (after state preparation with $U_S$).
The phase separating unitary $U_P(\gamma) = e^{-i\gamma H_C}$ then phases the amplitude of every basis state proportional to its energy and $\gamma$, keeping the same phase for basis states of same energy.
The mixing unitary $U_M(\beta) = \Id - (1-e^{-i\beta}) \ket{F}\bra{F}$ then deducts from all amplitudes of their input state $\ket{\psi}$ the same weighted average of all of its amplitudes, $((1-e^{-i\beta})/\sqrt{|F|}) \langle F \mid \psi \rangle$.
Therefore, basis states with the same energy are sampled with the same amplitude. 
This property holds for all feasible states (not only ground states) and is independent of the number of QAOA levels or the choice of angles for $U_M, U_P$. 
For a complete proof, see~\cite{baertschi2020grover}.

\subparagraph{Impact of noise:} While the original based Quantum Approximate Optimization Algorithm offers non-trivial provable performance guarantees already for a single for certain problems such as \textsc{MaxCut} on ($d$=3)-regular graphs~\cite{Farhi2014}, a higher number of rounds -- $p\geq 6$~\cite{wurtz2021qaoa} or even $p \in \Omega(\log(n)/d)$~\cite{bravyi2020qaoa} -- might be necessary to outperform the best-known classical approximation ratios.

However, simulations considering realistic noise models of current hardware show that performance in terms of approximation ratio may degrade even for $p=2$~\cite{qaoa-noise}. Noise leading to a decrease in performance has also been observed for growing problem sizes $n$, with the exception of problem graphs matching the hardware connectivity~\cite{harrigan2021quantum}. For the latter setting, experiments on actual hardware have shown an approximation ratio increase up to $p=3$ (mean) and $p=4$ (certain instances), but no further~\cite{harrigan2021quantum}. Furthermore, even with constanty improving hardware, complications may arise in the optimization of the variational angles $\beta,\gamma$ due to noise-induced barren plateaus~\cite{wang2020noise}.

All of these results refer to transverse field based QAOA, which does not have the property of sampling fairly among ground states. For Grover Mixer QAOA, we have a theoretical guarantee of fair sampling for ideal noise-less devices. At the same time we have a larger circuit depth of the mixer, asymptotically matching the circuit depth of the phase separator~\cite{baertschi2020grover}. Hence for current NISQ devices, we restrict ourselves to a single-round Grover Mixer QAOA, analysing the impact of noise on both ground state probability and fair sampling among ground states.

\subsection{Description of test circuits}
\begin{table}[ht!]
\makebox[\linewidth]{
\centering
    \begin{tabular}{|c|l|l|l|}
    	\hline
	Problem & Diagram & Ising Hamiltonian $H_C$ & Ground States $\ket{(q_0:=\,\uparrow)\ldots q_{n-1}}$ \\
        \hline
        (a) 
        & \begin{minipage}{2.75cm}
        \includegraphics[width=2.75cm, height=2cm]{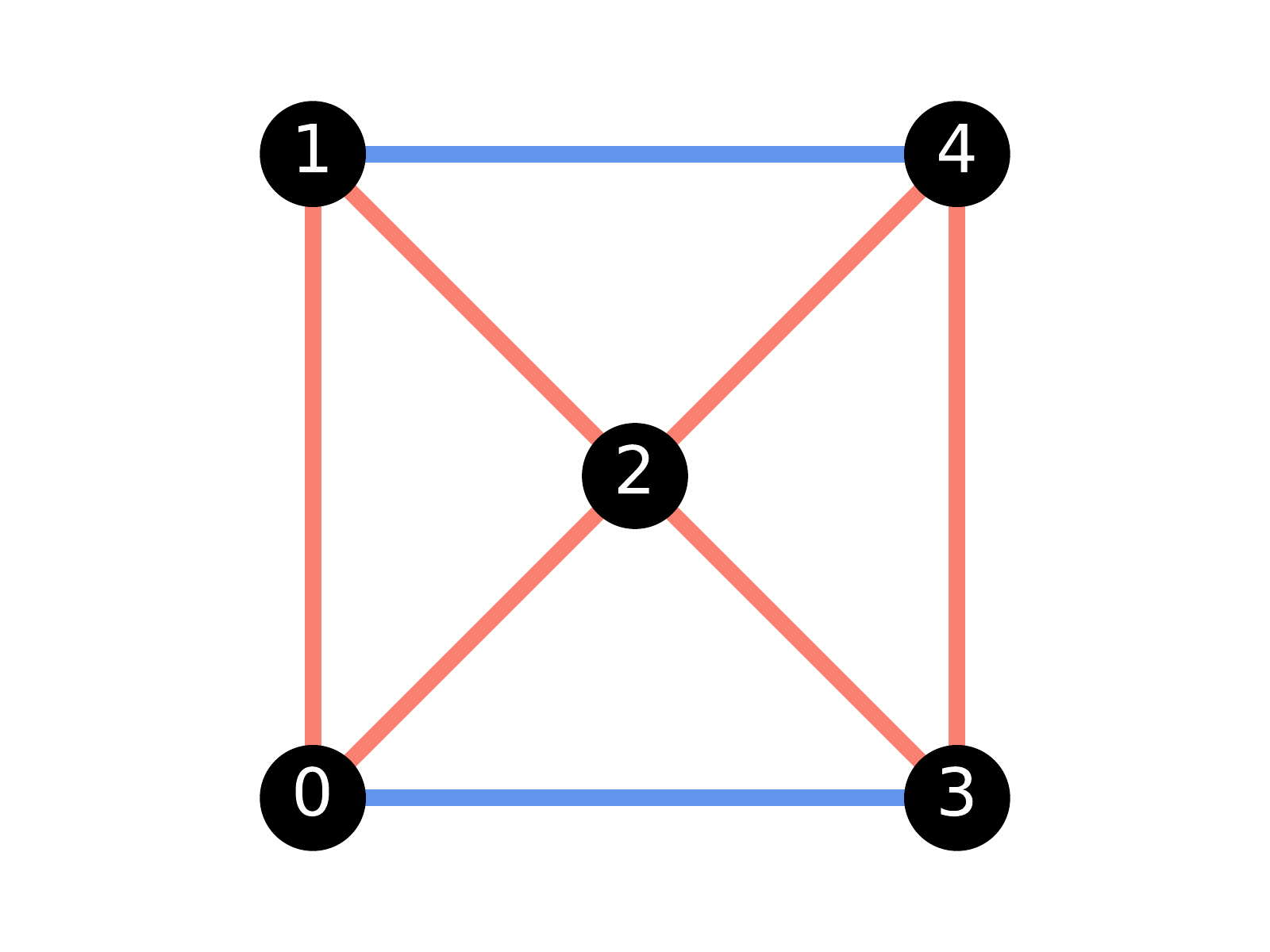}
        \end{minipage}
	& $\begin{aligned} & -[ Z_0(Z_1+Z_2-Z_3) \\&\qquad+Z_1(Z_2-Z_4)\\&\qquad+Z_2(Z_3+Z_4)+Z_3Z_4 ] \end{aligned}$
        & $\left|\uparrow \uparrow \uparrow \uparrow \uparrow \right\rangle ,\left|\uparrow \uparrow \uparrow \downarrow \downarrow \right\rangle ,\left|\uparrow \uparrow \downarrow \downarrow \downarrow \right\rangle$ \\[1cm]
        (b) 
        & \begin{minipage}{2.75cm}
        \includegraphics[width=2.75cm, height= 2cm]{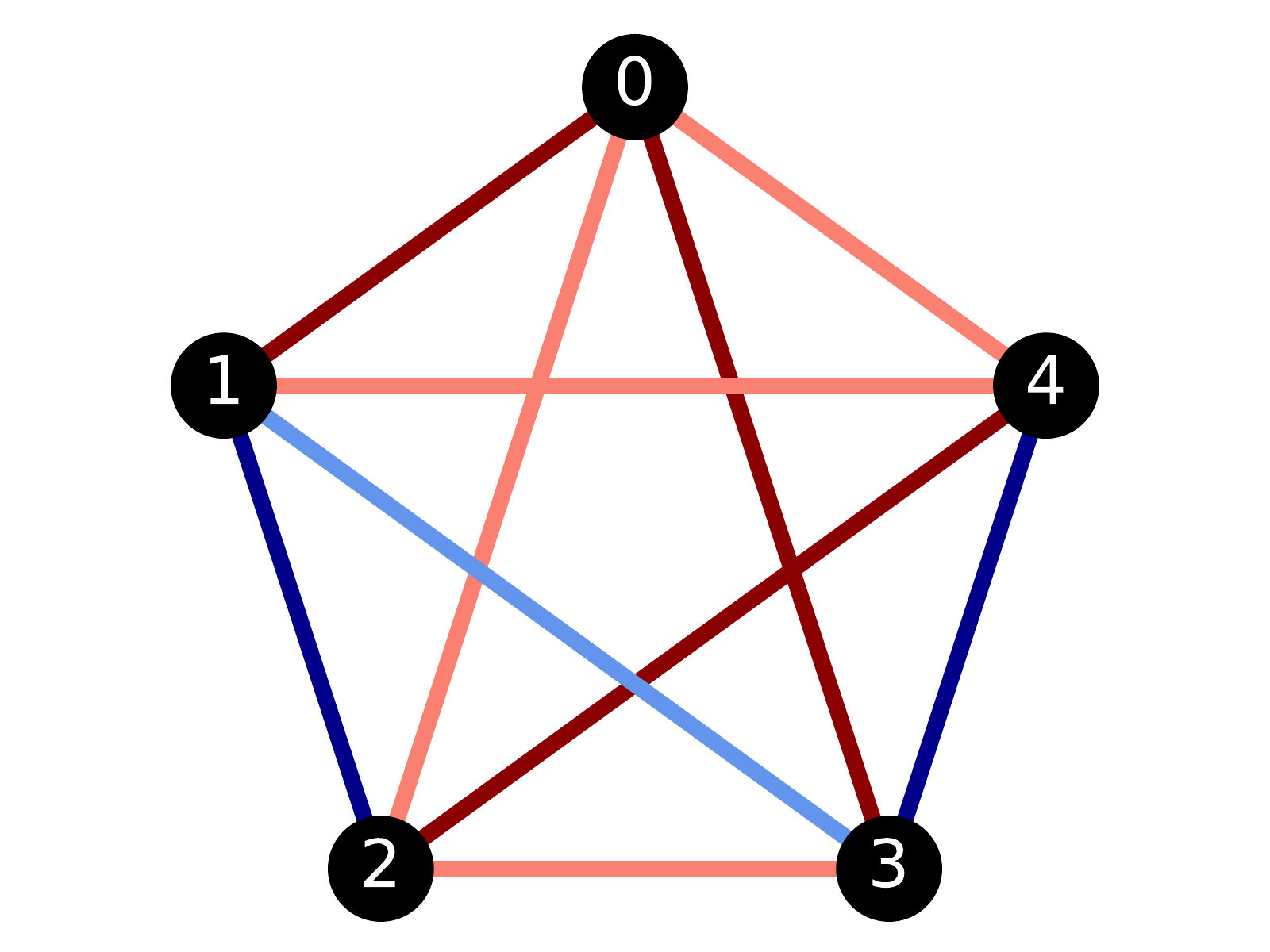}
        \end{minipage}
	& $\begin{aligned} & -[ Z_0 (2 Z_1 + Z_2 + 2 Z_3 + Z_4)\\&\qquad+Z_1 (-2 Z_2 - Z_3 + Z_4)\\&\qquad+Z_2(Z_3+2Z_4) - 2 Z_3 Z_4 ] \end{aligned}$
        & $\begin{aligned} &\left|\uparrow \uparrow \uparrow \uparrow \uparrow \right\rangle ,\left|\uparrow \uparrow \uparrow \downarrow \uparrow \right\rangle ,\left|\uparrow \uparrow \downarrow \uparrow \downarrow \right\rangle ,\\&\left|\uparrow \uparrow \downarrow \downarrow \uparrow \right\rangle ,\left|\uparrow \downarrow \uparrow \uparrow \uparrow \right\rangle ,\left|\uparrow \downarrow \uparrow \uparrow \downarrow \right\rangle \end{aligned}$  \\[1cm]
        (c) 
        & \begin{minipage}{2.75cm}
        \includegraphics[width=2.75cm, height= 2cm]{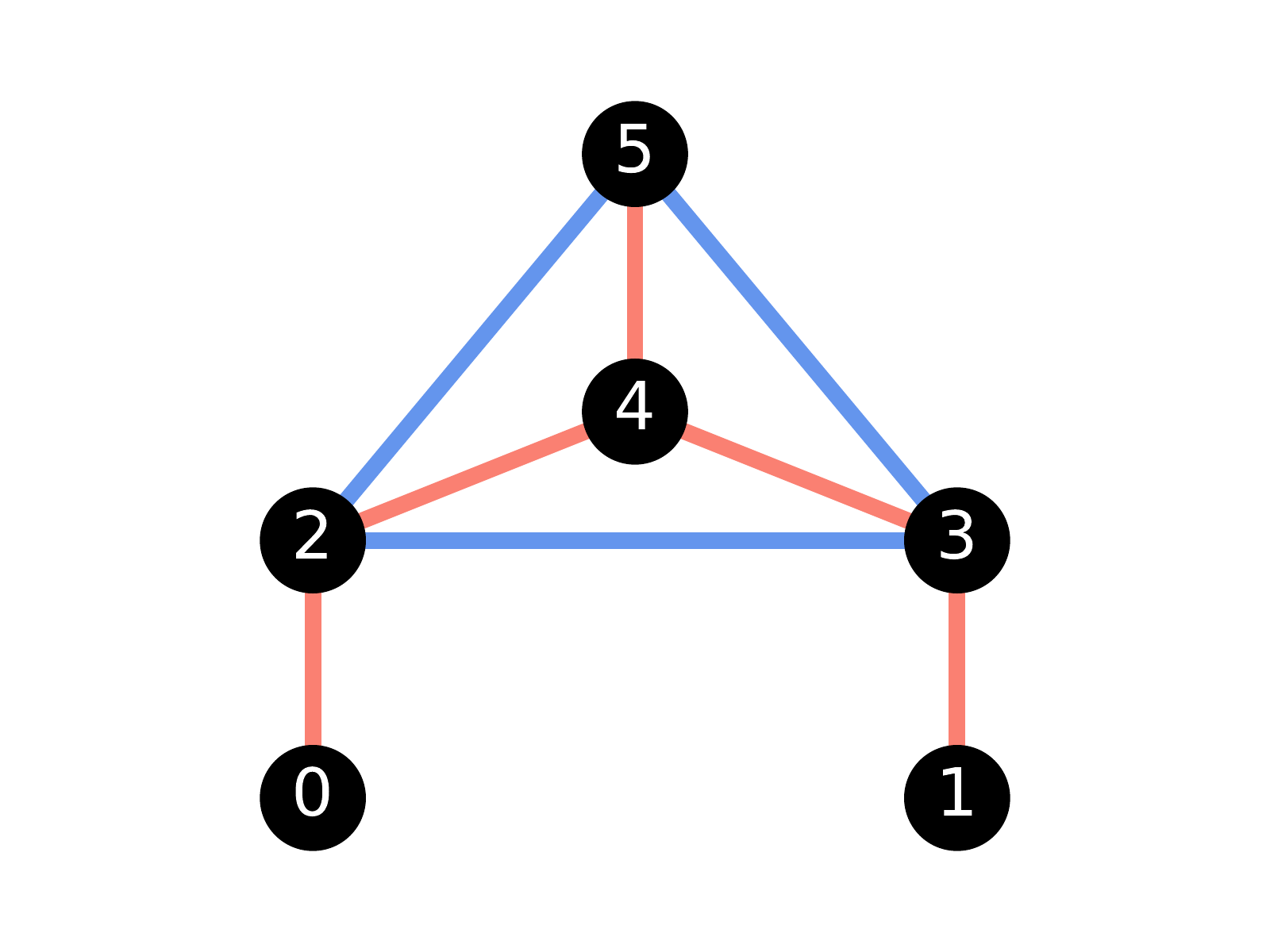}
        \end{minipage}
	& $ \begin{aligned}& -[ Z_0 Z_2 + Z_1 Z_3\\&\qquad + Z_2(-Z_3+Z_4-Z_5)\\&\qquad+Z_3(Z_4-Z_5)+Z_4Z_5 ] \end{aligned}$
        & $\left|\uparrow \uparrow \uparrow \uparrow \uparrow \downarrow \right\rangle ,\left|\uparrow \downarrow \uparrow \downarrow \uparrow \uparrow \right\rangle ,\left|\uparrow \downarrow \uparrow \downarrow \downarrow \downarrow \right\rangle$  \\[1cm]
        (d) 
        & \begin{minipage}{2.75cm}
        \includegraphics[width=2.75cm, height= 2cm]{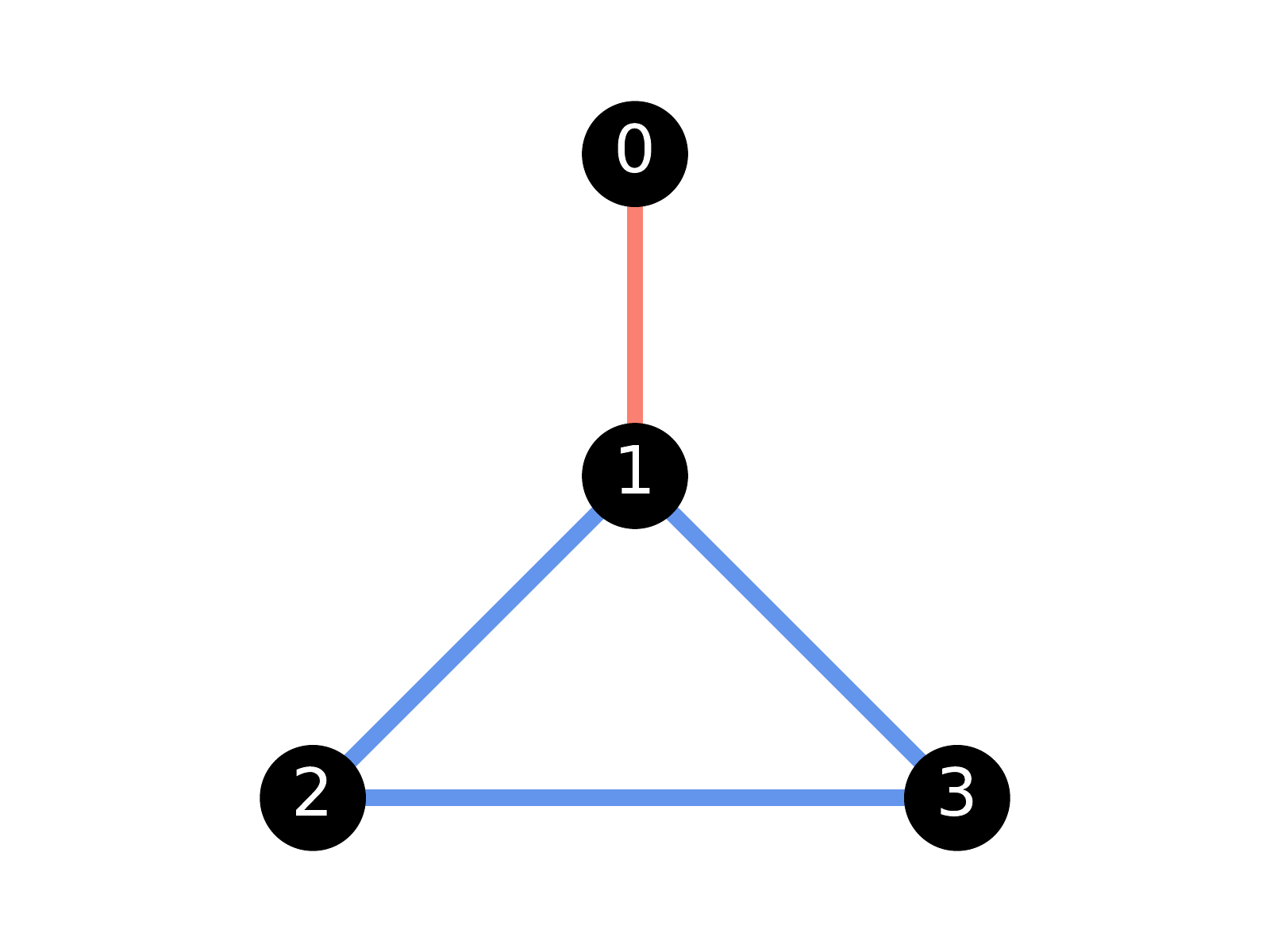}
        \end{minipage}
	& $ \begin{aligned} & -[ Z_0 Z_1\\&\qquad+Z_1(-Z_2-Z_3) - Z_2Z_3 ] \end{aligned}$
        & $\left|\uparrow \uparrow \uparrow \downarrow \right\rangle ,\left|\uparrow \uparrow \downarrow \uparrow \right\rangle ,\left|\uparrow \uparrow \downarrow \downarrow \right\rangle $  \\[1cm]
        (e) 
        & \begin{minipage}{2.75cm}
        \includegraphics[width=2.75cm, height= 2cm]{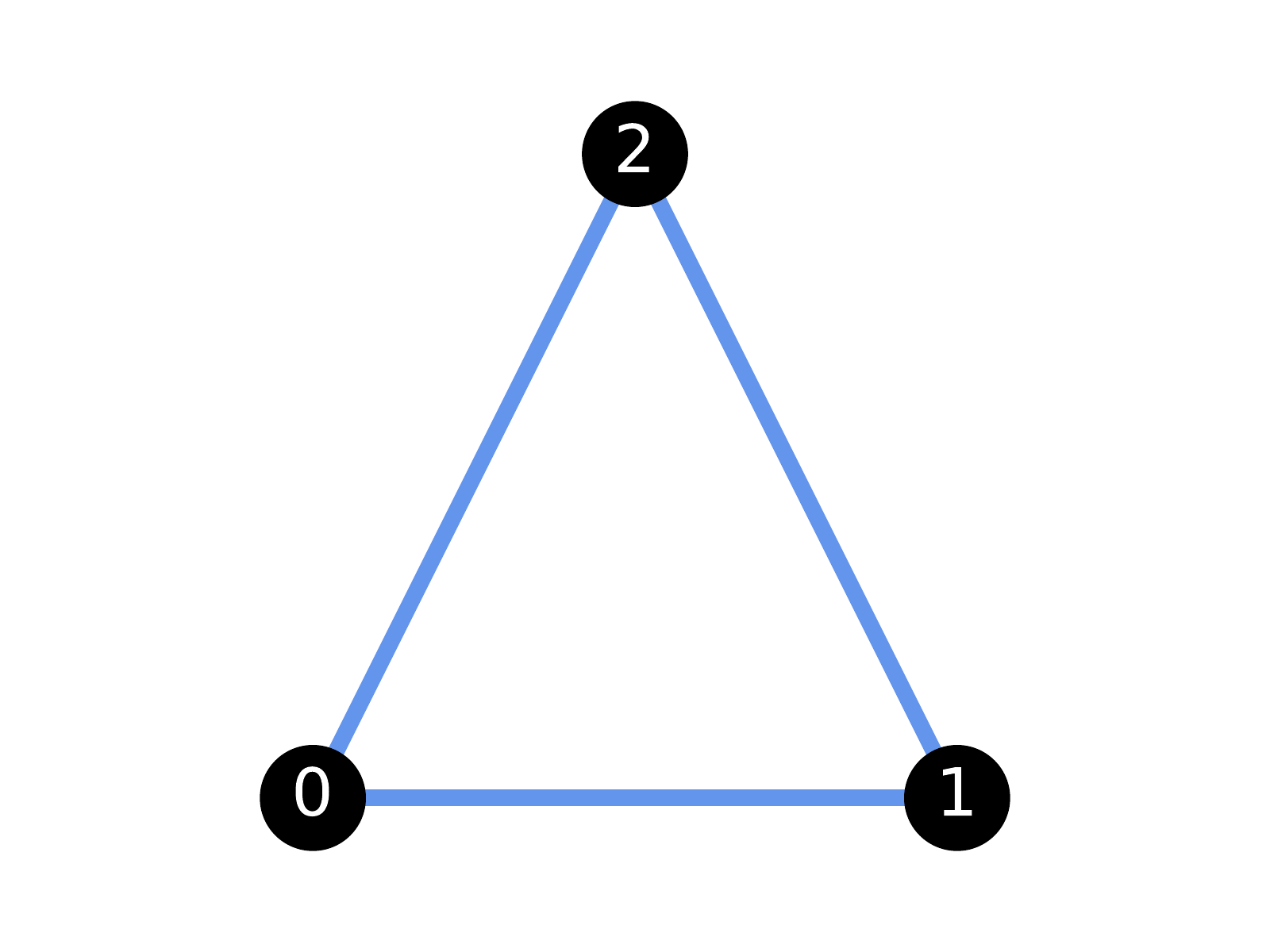}
        \end{minipage}
	& $-[ Z_0(-Z_1-Z_2)-Z_1Z_2 ] $
        & $\left|\uparrow \uparrow \downarrow \right\rangle ,\left|\uparrow \downarrow \uparrow \right\rangle ,\left|\uparrow \downarrow \downarrow \right\rangle$  \\[1cm]
        (f) 
        & \begin{minipage}{2.75cm}
        \includegraphics[width=2.75cm, height= 2cm]{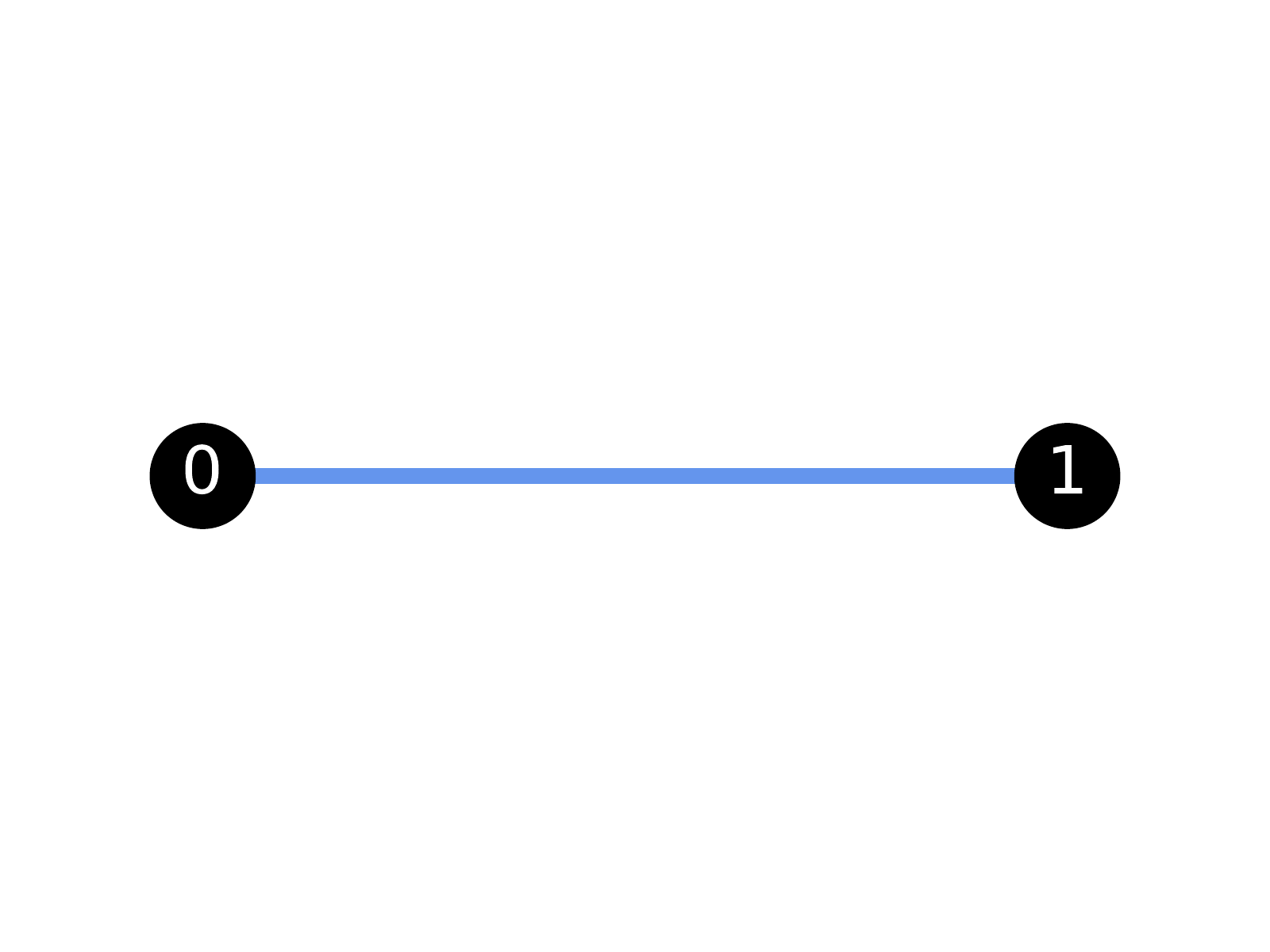}
        \end{minipage}
	& $-[ -Z_0Z_1 ] $
        & $\left|\uparrow \downarrow \right\rangle$ \\\hline
    \end{tabular}
    }
	\caption{Ising models with degenerate ground states to be studied on NISQ hardware.
	Problems (a)--(d) are from~\cite{K_nz_2019}. 
	All of the Ising models $H_C = -\sum J_{ij} Z_i Z_j$ have only quadratic terms and no linear terms.
	The dark red edges indicate a ferromagnetic $J_{ij} = +2$ coupling and the light red edges a ferromagnetic $J_{ij}=+1$ coupling;
	the light blue edges represent an antiferromagnetic $J_{ij} = -1$ coupling and the dark blue edges an antiferromagnetic $J_{ij}=-2$ coupling.
	Only ground states with $q_0=\,\uparrow$ are listed as all models are symmetric under simultaneous $\uparrow/\downarrow$-flips.
	Note that a symmetry-breaking fixed setting $q_0 := \, \uparrow$ results in a Hamiltonian $H_C'$ on qubits $q_1,\ldots,q_n$ (without $q_0$) with some linear terms. 
	}
	\label{tab:models}
\end{table}
In addition to the four problems from~\cite{K_nz_2019}, we additionally studied two simpler Ising models with degenerate ground states but only involving two and three qubits. 
While these models do not feature ground state suppression on quantum annealing hardware, we include them in this study because they represent the simplest non-trivial models with degenerate ground states and thus serve as useful baselines as compared to the more complicated models of~\cite{K_nz_2019}.
All of the models we studied are depicted in Table~\ref{tab:models}.

Our general procedure for generating test circuits was to begin with a 1-level Grover Mixer QAOA algorithm with $q_0$ fixed to $\uparrow$.
We chose a 1-level implementation of Grover-QAOA in order to keep circuit depth low. 
Following~\cite{K_nz_2019}, we fix $q_0:=\,\uparrow$ as all of the models in Table~\ref{tab:models} are $\uparrow/\downarrow$ symmetric.
This allows us to reduce the problems from Ising Hamiltonians $H_C$ with only quadratic terms acting on qubits $q_0,\ldots,q_{n-1}$ to new Hamiltonians $H_C'$ with some linear terms acting only on qubits $q_1,\ldots,q_{n-1}$. 
This also transforms $q_0$ into a classical control bit for the consecutive Grover Mixer, which can be removed. 
Thus we can embed Ising problems on $n$ qubits onto circuits with only $n-1$ qubits.
See Fig.~\ref{fig:qaoa} in Appendix~\ref{sec:circ-details} for a visual representation of a generic 1-level Grover Mixer QAOA for an unconstrained optimization problem, along with a description of the simplifications resulting from fixing $q_0$.

\begin{figure}[ht!]
	\centering
	\includegraphics[width=\linewidth]{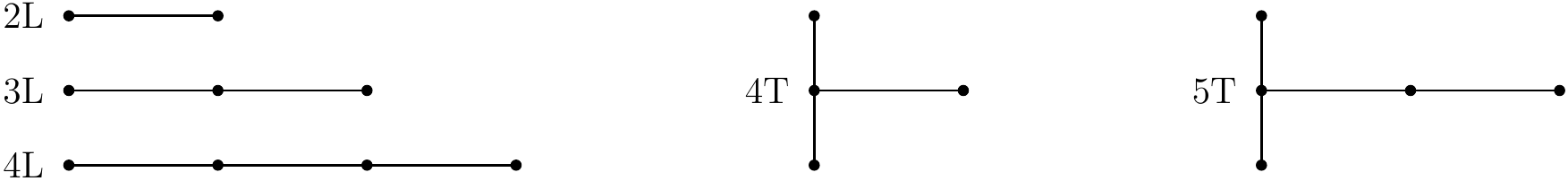}
	\caption{The hardware architectures we transpile our circuits to: Linear Nearest Neighbor connectivities on 2,3,4 qubits (2L,3L,4L), 
	as well as T-shaped connectivities on 4,5 qubits (4T,5T). These architectures form connectivity subgraphs of all IBM Q devices. 
	By fixing $q_0:=\,\uparrow$, each $n$-qubit problem in Table~\ref{tab:models} can be mapped to a $(n-1)$-qubit architecture.}
	\label{fig:architectures}
\end{figure}

We then compiled the circuits to match connectivity graphs and gates available on the IBM Q backend. 
Instead of employing the tools available in IBM's \verb|qiskit| software, we compiled the circuits by hand.
This had two benefits.
First, this reduced circuit depth by roughly a factor of two as compared to \verb|qiskit|.
Second, it allowed us to generate circuits for architectures that can be found as subgraphs of every hardware connectivity graph among the IBM Q backends, see Fig.~\ref{fig:architectures}.
Thus, the same exact circuit could be evaluated across many backends, and in many instances across many different qubits on the same backend.

For Problems (a) and (b), which involve 5 variables (and with $q_0:=\,\uparrow$ require only 4 qubits) we generated three distinct circuits, each on a different architecture: 4T, 4L, and 5T (see Fig.~\ref{fig:architectures} for notation). 
The 5T circuits for these problems employ an ancilla qubit, which allows us to discarded any sample in which the ancilla was not measured in the $\uparrow$ state.
The remaining Problems (c) - (f) each only have a single circuit associated with them.
Note that Problem (f), with the Hamiltonian $Z_0Z_1$ and ground states $\left|\uparrow \downarrow\right\rangle$ and $\left|\downarrow \uparrow\right\rangle$, is uniquely simple.
In this case we do not fix $q_0:=\,\uparrow$, nor do we employ Grover-QAOA, as the problem can be ``solved'' with an almost-trivial circuit of a Hadamard on qubit $q_0$, a $X$-gate on qubit $q_1$, and a consecutive CNOT with control $q_0$ and target $q_1$.  
Due to the small number of qubits and gates, we use this circuit to explore the highest reaches of accuracy and fairness capable with the hardware. 

See Table~\ref{tab:parameters} for a brief description of all of the circuits, including their architectures, gate count, and $\bbeta$, $\bgamma$ parameters. 
We describe our compilation technique in detail in Appendix~\ref{sec:circ-details}, and give QASM representations of our circuits in the ancillary file associated to this note.

\section{Methodology}

We evaluated the previously described set of circuits across a wide swath of hardware on the IBM Q platform.
Furthermore, we collected data in two large sets, one in October 2020 and the other in August 2021, allowing for an analysis of the change in fairness of the IBM Q platform over time. 
Since our interest was not in identifying the best performing qubits, but rather observing trends in circuit correctness vs. fairness, we evaluated the same circuit many times per chip, each time using a different subset of qubits.
For example, testing the circuit for Problem (e), which has only two qubits, on a chip with 5 qubits in an LNN architecture would involve 4 separate evaluations (on the pairs of qubits $\{q_i,q_{i+1}\}$ for $i=1\ldots4$).
The machines we tested, along with the number of qubit subsets for each architecture, are listed in Table~\ref{tab:backends}.
It is important to emphasize that there was no additional transpilation or optimization of these cicuits beyond that discussed in the previous section. 

We collected 40960 shots for each circuit on each set of applicable qubits, which gave us sufficient data to achieve repeatable statistics.  
None of the hardware consistently reported fair results, i.e. hardware noise introduced significant biases for even the simplest circuits.
In order to characterize how fair a given set of qubits performed with a given circuit, we adopted the metric: how many shots does one need of a given circuit on a given set of qubits in order to \emph{reject} the hypothesis that these qubits are fair samplers at 95\% significane level?
This is equivalent to the question of: how many flips (on average) does it take to show that a coin is unfair? 
A 90/10 coin will take relatively few flips to reveal its bias, whereas a 51/49 coin will take many flips. 
We call this metric ``number of shots to reject fair sampling.''

To calculate the number of shots to reject fair sampling, we utilized Pearson's $\chi^2$ test, which evaluates how likely a set of observations $\{O_i\}$ is given some expected null hypothesis values $\{E_i\}$. 
If a given circuit has $d$ degenerate ground states, and $n_{g.s.}$ is the number of ground state observations (out of the 40960 total shots), our null hypothesis is that
\begin{equation}
 	E_i = E = \frac{n_{g.s.}}{d}.
\end{equation} 
The $\chi^2$ value is determined via
\begin{equation}
	\chi^2 = \sum_i\frac{(O_i-E)^2}{E}.
\end{equation} 
The null hypothesis is rejected at a significance level based on the value of $\chi^2$ as compared against the upper-tail critical values of the $\chi^2$ distribution 
\begin{equation}
	f(x,k) = \frac{x^{k/2-1}e^{-x/2}}{2^{k/2}\Gamma(k/2)},
\end{equation}
where $k$ is the number of degrees of freedom in the observed data. 
For example, with five degrees of freedom a $\chi^2$ value of $11.070$ indicates a rejection of the null hypothesis with $95\%$ significance level. 

\begin{table}[t!]
	\centering
	\begin{tabular}{|c|c|c|c|r|c|c|}
		\hline
		Prob.	& Arch. & \# Rot. & \# CNOT & ($\beta$, $\gamma$)\qquad~\phantom{.}	& $\bgstateT H_C \bgstate$	& GSP \\
		\hline
		(a)	& 4L & 57 & 35 & $(-1/2,-11/12) \cdot \pi$	& -2.682 out of -4	& 0.498			\\
		(a)	& 4T & 51 & 29 & $(-1/2,-11/12) \cdot \pi$	& -2.682 out of -4	& 0.498			\\
		(a)	& 5T & 47 & 25 & $(-1/2,-11/12) \cdot \pi$	& -2.682 out of -4	& 0.498			\\
		(b)	& 4L & 59 & 39 & $(-11/15,-17/60) \cdot \pi$	& -4.228 out of -5	& 0.846			\\
		(b)	& 4T & 53 & 32 & $(-11/15,-17/60) \cdot \pi$	& -4.228 out of -5	& 0.846			\\
		(b)	& 5T & 49 & 29 & $(-11/15,-17/60) \cdot \pi$	& -4.228 out of -5	& 0.846			\\
		(c)	& 5T & 90 & 62 & $(-23/60,1/15) \cdot \pi$	& -1.563 out of -4	& 0.215			\\
		(d)	& 3L & 26 & 14 & $(-5/12,1/10) \cdot \pi$	& -1.319 out of -2	& 0.702			\\
		(e)	& 2L & 16 & 4 & $(-23/60,3/5) \cdot \pi$	& -0.999 out of -1	& 1.000			\\
		(f)	& 2L & 2 & 1 & N/A\qquad~\phantom{.}	& N/A & 1.000			\\
		\hline
	\end{tabular}
	\caption{Details for 1-round Grover-QAOA fair sampling circuits studied on NISQ hardware. For each circuit, we list the problem from Table~\ref{tab:models} being solved, the qubit connectivity from Fig.~\ref{fig:architectures} employed, the total number of single-qubit (i.e.~rotation) gates, the total number of CNOT gates, the $\bbeta, \bgamma$ parameters used, the expectation value $\bgstateT H_C \bgstate$, 	and the probability of selecting a ground state (GSP).}
	\label{tab:parameters}
\end{table}

In principle, the value of $\chi^2$ is itself a measure of fairness for a set of observational data, however we found it a poor metric for our application. 
This is because the actual significance of a given $\chi^2$ value is dependent on the size of the data set as well as the number of degrees of freedom.
Since our circuits reported widely varying numbers of ground state observations, and had different numbers of degenerate ground states, comparing $\chi^2$ values across circuits did not compare apples to apples.
One could instead report the significance level at which the observational data rejects the fair sampling hypothesis, as that takes sample size and degrees of freedom in to account.
However, current hardware rejects the fair sampling hypothesis at significance levels so close to 100\% that numerical differences between rejection levels were difficult to meaningfully interpret.
Instead, determining how many samples from a distribution are necessary (on average) to reject fair sampling at 95\% significance can be done with observational data of any size and degrees of freedom, and provides an intuitive measure of the fairness of the underlying qubits. 

A formal description of our algorithm for determining the number of shots to reject fair sampling can be found in Appendix~\ref{sec:NSRFS}, here we describe the method in general terms.
Based on our observational data of a given circuit $C$ on a set of qubits $Q$, we determined the relative frequency of observing each ground state.
We then generated 1000 synthetic samples, each containing $n$ ground state observations matching the relative frequency in the observed data. 
For each sample we calculated $\chi^2$, then took the median value for the set of 1000 samples\footnote{We found that taking fewer than 1000 samples resulted in inconsistent median values, and more then 1000 samples resulted in slow performance.}.
We repeated for increasing values of $n$ until we arrived at a median value of $\chi^2$ matching the level necessary to reject the fair sampling hypothesis with $95\%$ statistical signifance given the number of degenerate ground states.
As an example, with this methodology a coin biased 60/40 requires an average of 74 flips to reject fair sampling.

\section{Results}
As mentioned in the introduction, none of the qubits or circuits consistently reported fair sampling. 
Instead, we sought to understand the correlation between qubit fidelity and fair sampling. 
Our method for quantifying the degree of fair sampling for a circuit on a piece of hardware was described in the previous section. 
We used two complementary methods of quantifying qubit fidelity. 
The first is simply the frequency with which a given set of qubits accurately identify an optimal solution for the problem at hand.
In this context, higher fidelity qubits are those that obtain a ground state more frequently. 
The benefit of this method is that it relies on direct observation of qubits solving a real problem, while the downside is that it can't be used to directly compare results from different problems.
Our second method for quantifying qubit fidelity is to use the hardware error rates as reported by \verb|qiskit|.
This method, which we will discuss in more detail later in this section, has the benefit of more accurately comparing results from different circuits.
However, the practical relevance of this reported error data for our application is unclear.

A broad issue for experimental tests of existing NISQ hardware is the highly stochastic nature of the results. 
We observed significant variance in all of our observed data, and therefore used a very large experiment count (with 7440 distinct evaluations of the IBM Q hardware) to observe consistent trends. 
To be precise, each data point in the following plots is generated by specifying: a circuit to be evaluated, an IBM Q backend, and a subset of qubits on the backend which match the connectivity of the given circuit. 
We then collected 40960 shots of the circuit on those specific qubits, and the results of those shots correspond to a single data point in the plots below. 
This large number of data points allowed us to identify repeatable positive or negative correlations, and we include lines of best fit to indicate these trends.

\subsection{Ground state probability vs. fairness}
\begin{figure}[ht!]
	\includegraphics[width=\textwidth]{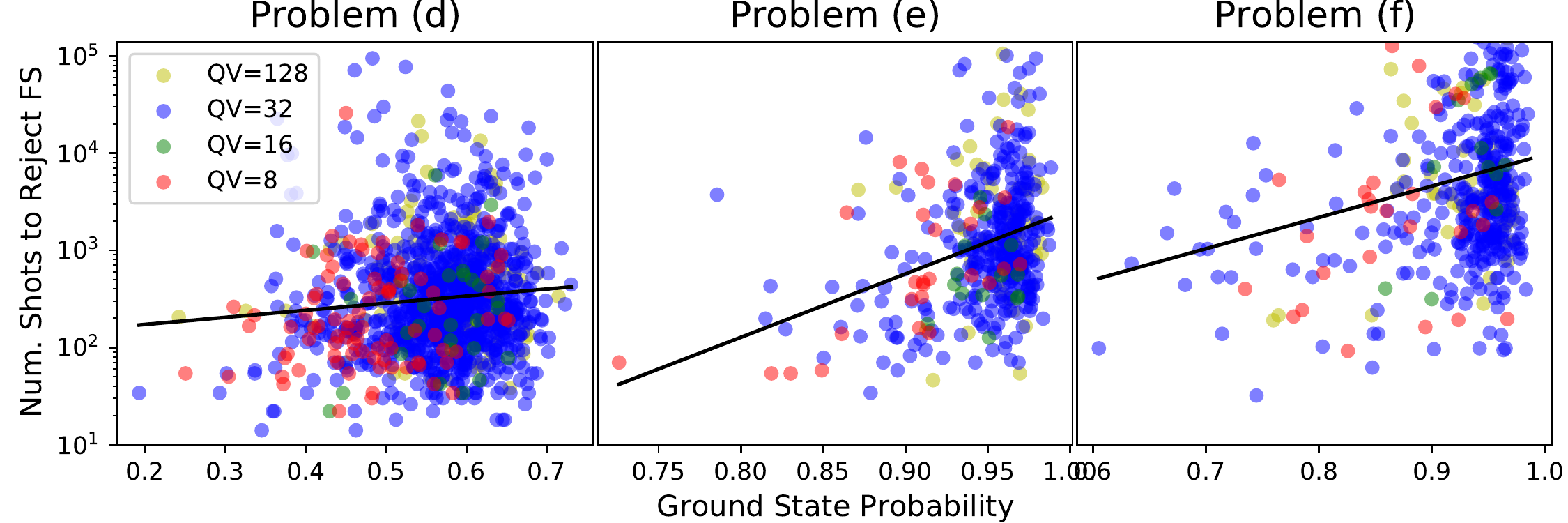}
	\caption{Each data point represents 40960 shots of the specified circuit on a fixed set of qubits on an IBM Q backend. The Quantum Volume (QV) of the backend is indicated by yellow for QV128, blue for QV32, green for QV16, and red for QV8. In these shorter circuits we observe a positive correlation (as indicated by the black lines of best fit) between ground state probability and number of shots to reject fair sampling.}
	\label{fig:soft}
\end{figure}

With that proviso out of the way, let us now discuss the results of directly comparing quantity (as measured by the number of ground state observations) vs. quality (as measured by the number of shots to reject fair sampling).
We observed a positive correlation between ground state probability and fair sampling for three circuits, those solving the ``easy'' problems: (d), (e), (f), see Fig.~\ref{fig:soft}.
Note that these circuits all have $\le 3$ qubits and circuit depths 40, 20, and 3 respectively. 
We call this ``soft'' unfairness, in that improving ground state probability and fairness of sampling are both accomplished by evaluating the circuit on the highest performing qubits possible. 

\begin{figure}[ht!]
	\includegraphics[width=\textwidth]{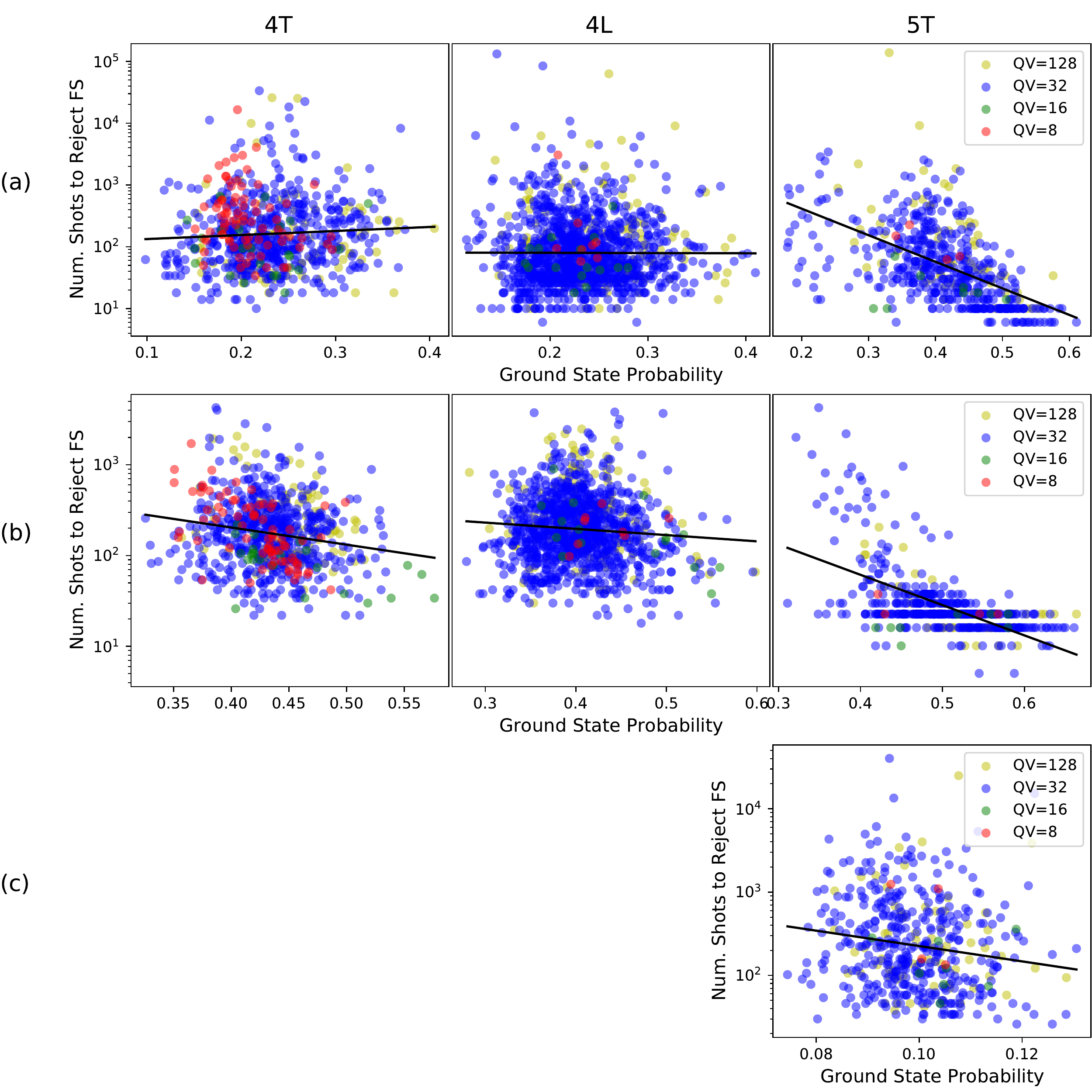}
	\caption{In these longer circuits, involving four or five qubits and between 72 and 152 gates, we mostly observe a negative correlation between ground state probability and number of shots to reject fair sampling, with Problem (a) on the 4L and 4T architectures as the exceptions.}
	\label{fig:hard}
\end{figure}

With the remaining circuits we observed more complicated behavior.
For Problem (a), the 4T circuit exhibits ``soft'' unfairness, the 4L circuit shows no variation in fairness with respect to ground state probability, and the 5T circuit has a \emph{negative} correlation between ground state probability and fairness, see Fig.~\ref{fig:hard}.
In other words, qubits that more frequently obtained ground states did so with increasing bias.
All circuits for Problems (b) and (c) share this negative correlation.
We call this ``hard'' unfairness, as improving ground state probability results in less fair results, and vice versa. 

We believe this ``hard'' unfairness comes from the fact that long circuits on noisy qubits produce increasingly random results.
This means that the likelihood of selecting a ground state decreases, but the distribution across ground states is fairly even. 
To test this conjecture, we compared the fairness and ground state probabilities of both a ``soft'' and ``hard'' circuit as functions of the error rate of the qubits.

\subsection{Aggregate error rate vs. fairness}
The IBM Q hardware reports the latest calibration data for a given chip at the time of evaluating a circuit. 
Specifically, \verb|qiskit| reports the error rates for their native gateset (CX, SX, RX, and RZ) as well as individual qubit measurement error.
We used this data to generate what we call an aggregate error of the chip and circuit in combination. 
For a circuit $C$ acting on qubits $q_1,\ldots,q_l$, composed of a sequence of gates $g_1,\ldots,g_k$  evaluated on a chip $H$, this aggregate error $E_{C,H}$ is calculated via
\begin{equation}\label{eq:aggregate-error}
	E_{C,H} = 1-\left(\prod_{i=1}^k(1-e_i)\right)\left(\prod_{i=1}^l(1-m_i)\right),
\end{equation}
where $e_i$ is the error for gate $g_i$, and $m_i$ is the measurement error for qubit $q_i$, as reported by \verb|qiskit| at the time of evaluation. 
We believe this aggregate error metric evaluates roughly the likelihood of success of a given circuit on a specific piece of hardware.
However, the usefulness of the error rates as reported by \verb|qiskit| has been drawn in to question~\cite{wilson2020justintime,proctor2020measuring}. 
These experiments therefore serve as a test of the relevance of this reported error value in the context of fair sampling. 

\begin{figure}[t!]
	\includegraphics[width=\textwidth]{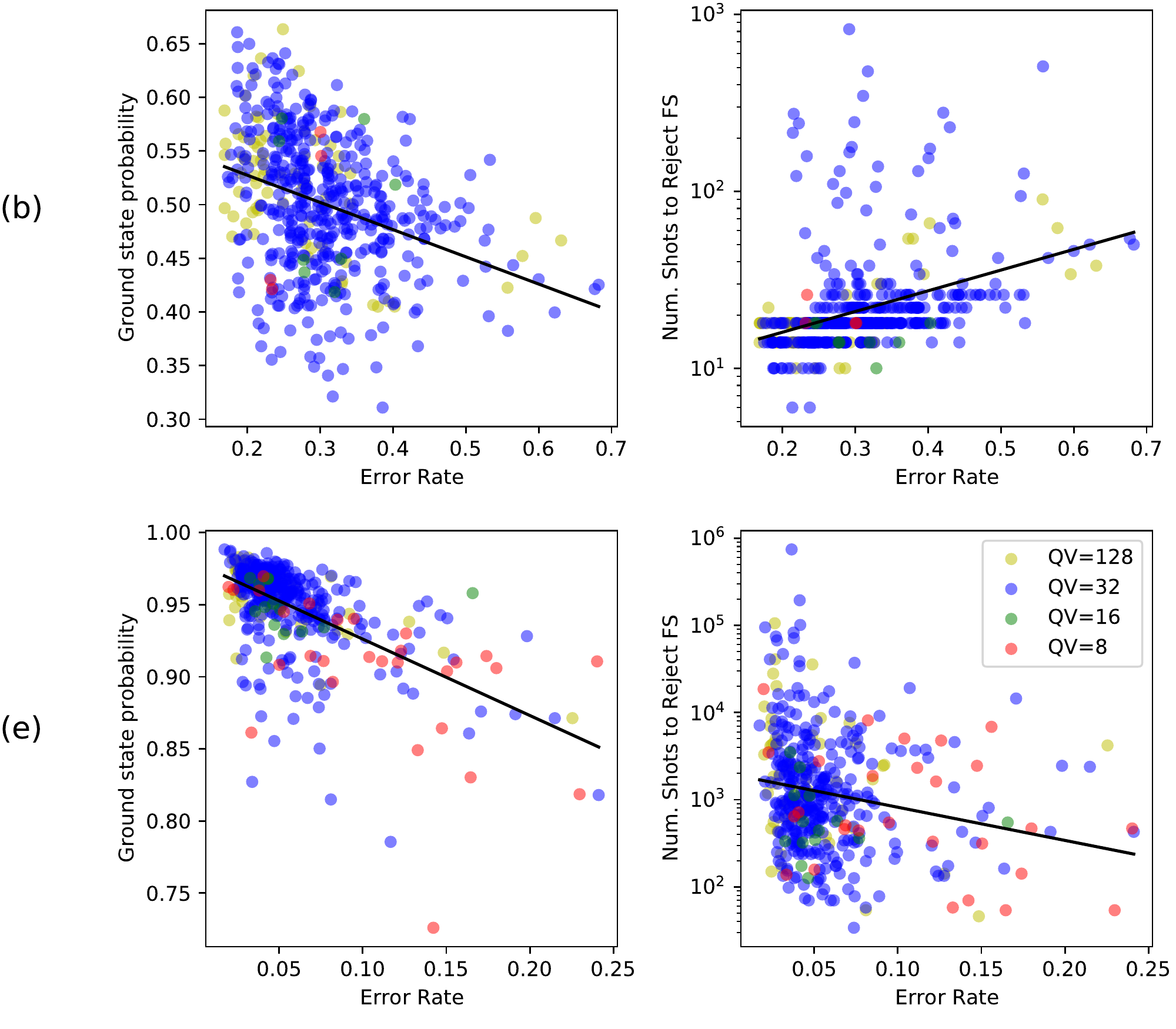}
	\caption{Results for a hard circuit, Problem (b) on the 5T architecture, as compared to results for a soft circuit, Problem (e). 
	All circuits studied featured a negative correlation between ground state probability and aggregate error, as pictured here for these two circuits in the leftmost plots. 
	For hard circuits, increase in aggregate error leads to behavior reminiscent of a random number generator, decreasing ground state probability but increasing fairness, as indicated in the top right plot.}
	\label{fig:soft-v-hard}
\end{figure}

In Fig.~\ref{fig:soft-v-hard} we see that both the soft and hard circuits report lower ground state probabilities as aggregate error rates increase, as expected\footnote{In fact, we observed a decrease in ground state probability as aggregrate error increased for all circuits, which serves as a sanity check on our methodology.}.
For Problem (e), we see that increased error rates results in less fair results. 
In other words, the errors result in systemic bias in favor of some ground states over others.
For Problem (b) on the 5T architecture, we see the opposite behavior: increased error rates result in fairer solutions. 
We can see in Fig.~\ref{fig:soft-v-hard} that the likelihood of obtaining a ground state is $\approx 0.4$ in the high error rate case. 
There are 6 degenerate ground states for this problem, out of 16 total states. 
Therefore a circuit that selected a state at random would select a ground state with $6/16 = 0.375$ probability. 

\begin{figure}[t!]
	\includegraphics[width=\textwidth]{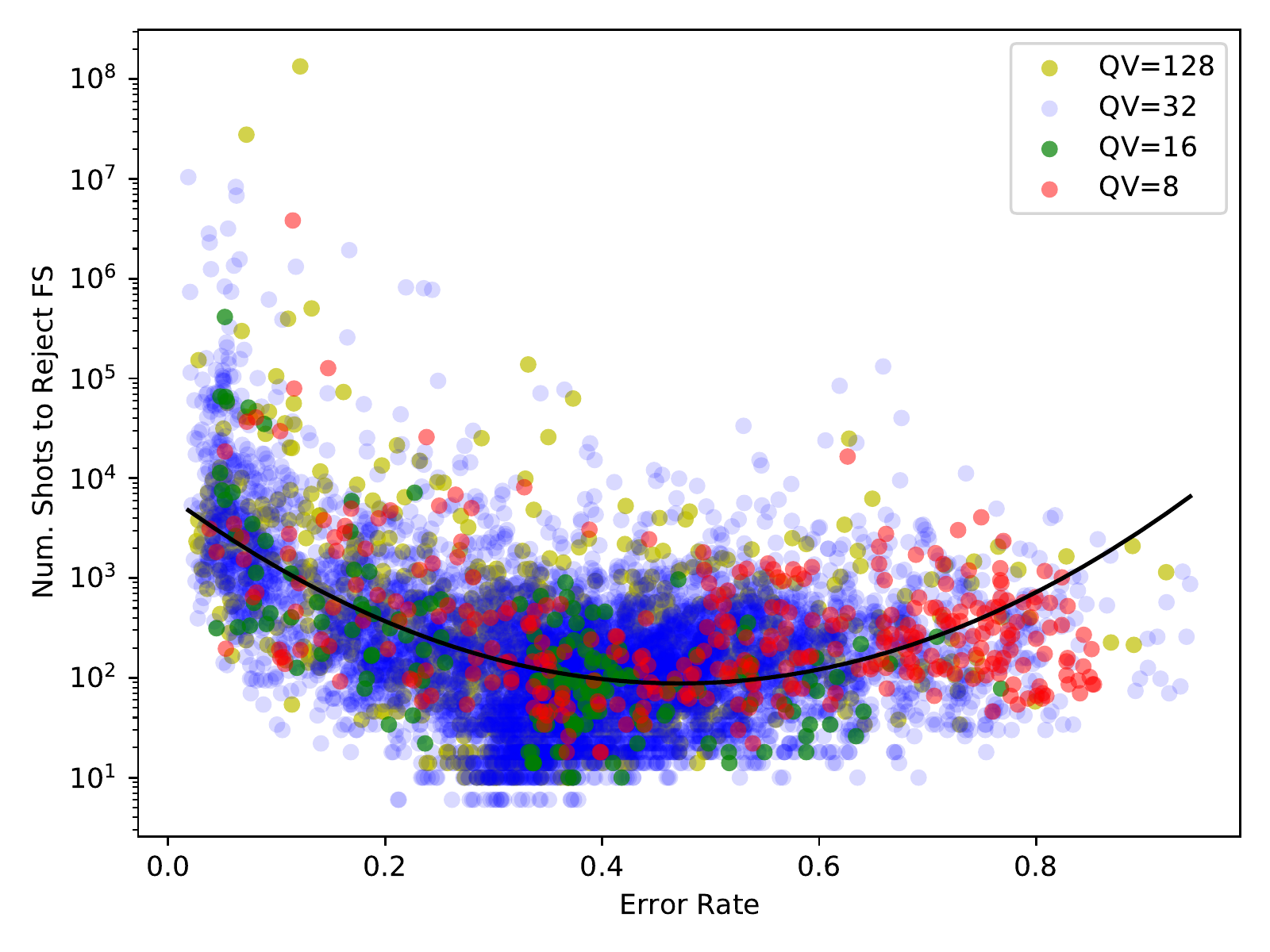}
	\caption{Fairness as a function of aggregate error rate across all experiments. Fairness is highest for small and large error rates, with a decrease in the middle. This indicates that structured errors dominate in the middle regime, while unstructured errors play an increasingly important role as error increases.
	The log-quadratic line of best fit (in black) is included to graphically indicate these overall trends.}
	\label{fig:error-v-fairness}
\end{figure}

By combining all of the data in to a single plot we can get a bigger picture view of how structured vs. unstructured errors impact our results. 
In Fig.~\ref{fig:error-v-fairness} we see that fairness is highest for low aggregate error, decreases as the error approaches approximately 0.5, and then begins increasing again.
This suggests that structured, i.e. biased, errors dominate in the middle of our error spectrum. 
As the overall error rate increases, either due to longer circuits or worse performing qubits, the net result is more unstructured error and thus fairer results. 

\subsection{Comments on the IBM Q system performance}

\begin{figure}[t]
	\includegraphics[width=\textwidth]{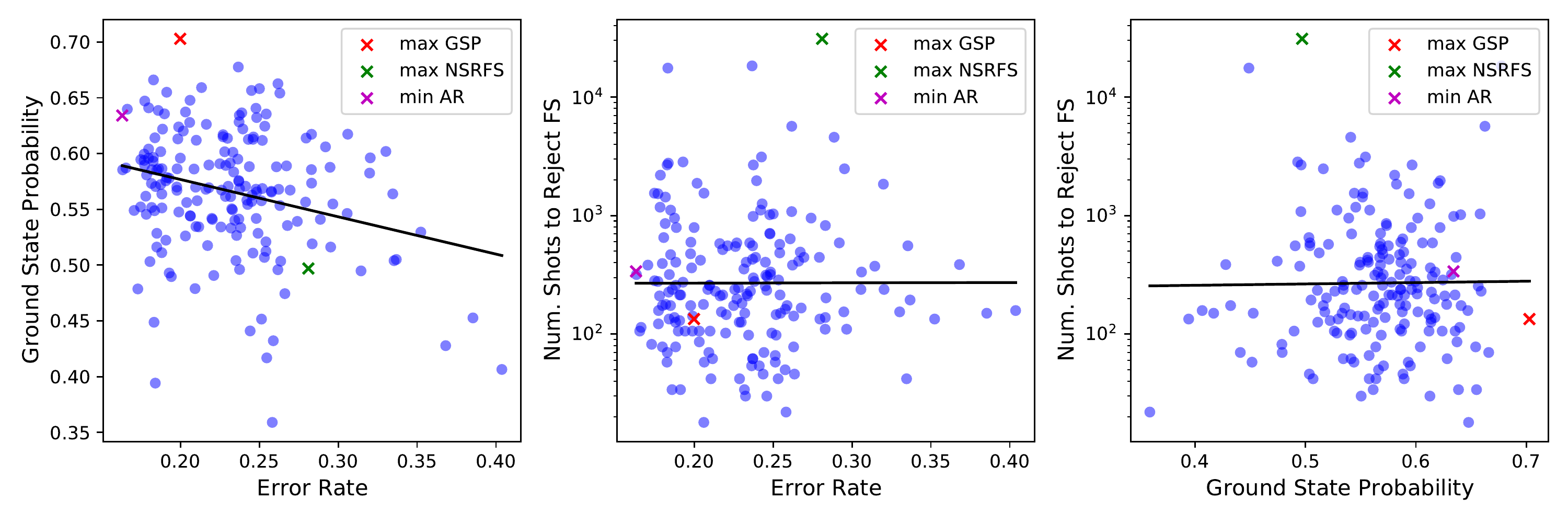}
	\cprotect\caption{Detailed results for Problem (d) evaluated on all 3-qubit subsets of the \verb|manhattan| backend. 
	Each plot represents the same underlying data, and the set of qubits with the maximum ground state probability (max GSP), maximum number of shots to reject fair sampling (max NSRFS), and minimum aggregate error (min AR) are highlighted. 
	This plot shows the high variance in results from a single backend.}
	\label{fig:prob-d-manhattan}
\end{figure}

We believe these results constitute a serious evaluation of the IBM Q system. 
Our two most prominent observations based on this data are:
\begin{description}
	\item[Correlation between Ground State Probability and Fairness shifted.]
	By specifically comparing the data generated in October 2020 against the data generated in August 2021, we can see how the overall fairness of the IBM Q system has evolved. 
	Interestingly, if one only looks at the 2020 data, all circuits from Problem (a) exhibit ``hard'' unfairness. 
	Running the experiments again in 2021, the 4T circuit shifted into the ``soft'' regime, see Fig.~\ref{fig:prob-a-improv}. 
	Furthermore, all circuits saw the slope of the fairness vs. ground state probability trend lines increase.
	Meanwhile, the average ground state probability across all problems decreased slightly, from $44.7\%$ to $43.4\%$. 
	The average fairness increased significantly in the new data, however this is due solely to a few data points in the two-qubit problems with extremely high ($\mathcal{O}(10^8)$) number of shots to reject fair sampling.
	The median number of shots to reject fair sampling actually decreased when looking at the new data, from 190 to 142. 
	From this data we conclude that the newer machines are more dominated by structured error as opposed to random error.
	\begin{figure}[t]
	\includegraphics[width=\textwidth]{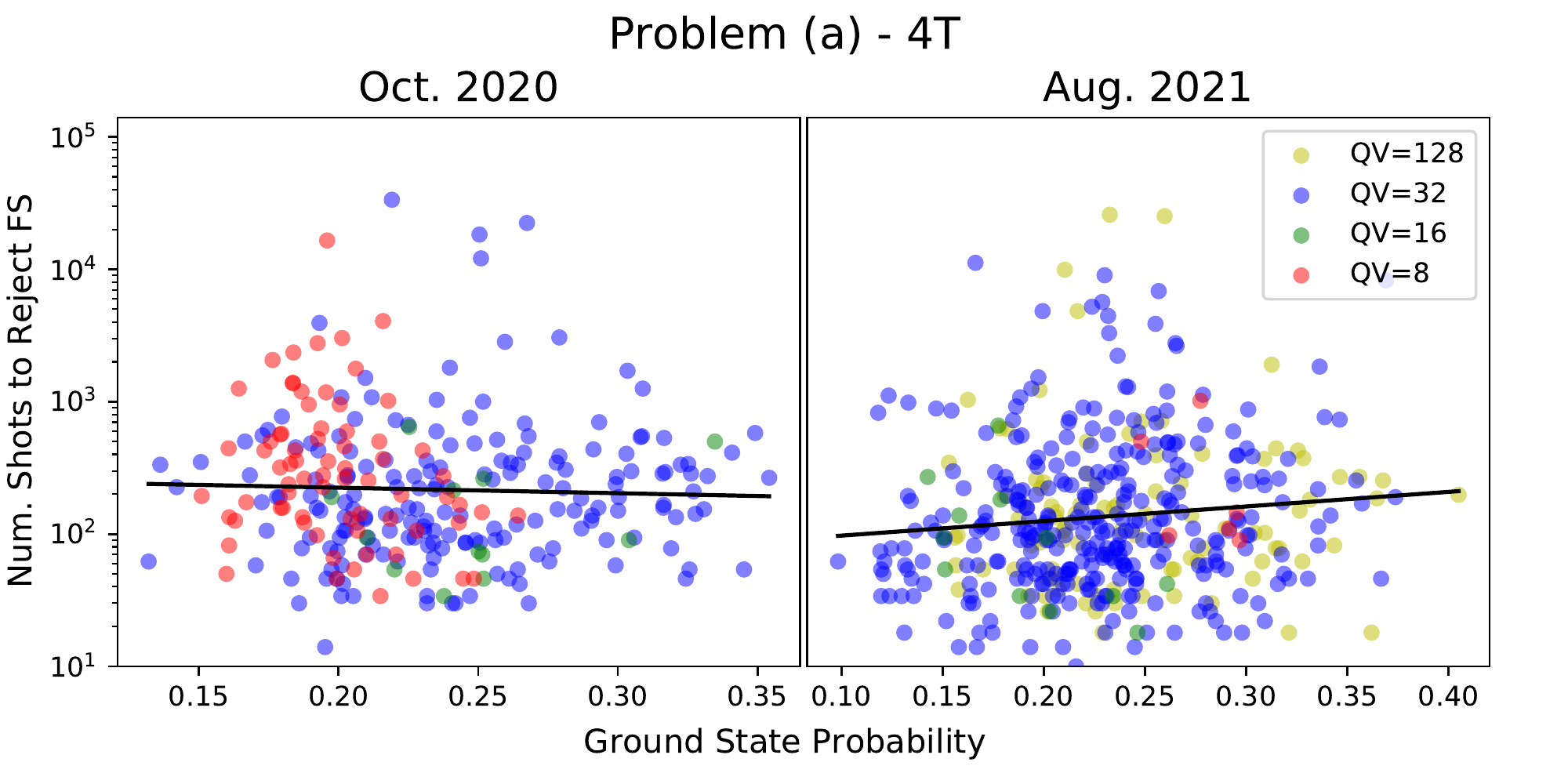}
	\cprotect\caption{Results for Problem (a) on the 4T architecture. 
	Data collected across the IBM Q system in October 2020 showed a negative correlation between ground state probability and number of shots to reject fair sampling, whereas data collected in August 2021 showed a positive correlation.
	This indicates that the newer dates are more dominated by structured errors.}
	\label{fig:prob-a-improv}
	\end{figure}

	\item[Quantum Volume only loosely correlates with fairness.]
	The IBM metric of Quantum Volume~\cite{Cross_2019} did broadly correlate with decreased aggregate error, which led to improved fairness in short circuits (see Fig.~\ref{fig:soft-v-hard}).
	However, the Quantum Volume metric is only based off of the highly tuned performance of the best performing qubits available on a machine, instead of the overall hardware performance. 
	This study, which ranges over all available qubits on a wide range of available chips, gives a more comprehensive view of performance.
	For example, we have plotted the results for Problem (d) evaluated on all 190 connected 3-qubit subsets on the \verb|manhattan| backend (which has QV32 and 65 qubits), see Fig.~\ref{fig:prob-d-manhattan}.
	In these plots one can see a wide range of results, and the best performing qubits for a single metric (ground state probability, number of samples to reject fair sampling, and aggregate error rate) can perform quite poorly when evaluated by a different metric. 
	Thus, this study emphasizes the need for more comprehensive benchmarking that does not only focus on random circuits evaluated on a small subset of qubits, as also discussed in~\cite{proctor2020measuring}.

\end{description}

\section{Measurement Error Mitigation}\label{sec:mitigating-discussion}
In this section we analyze how measurement error affects fairness. 
The measurement error included in our aggregate error metric (see eq.~\ref{eq:aggregate-error}) is for each individual qubit.
This tensored approach paints an incomplete picture of measurement error, as there are correlations between qubits leading to correlated errors~\cite{tannu2019mitigating}. 
Therefore, we collected more thorough measurement error data immediately before evaluating our fair sampling circuits. 
Specifically, for a given $n$-qubit circuit we explicitly prepared each of the $2^n$ basis states and then measured the likelihood that the all of the other states would be observed.
Using this data, we created a measurement error calibration matrix for each set of qubits and used the inverse of that matrix to correct for measurement error~\cite{qiskit-me-mitigation-matrix}.
Due to the small number of qubits in our problems under consideration, this simple approach is feasible; we relied on it rather than more advanced measurement error mitigation schemes~\cite{Funcke:2020olv,tannu2019mitigating,qiskit-me-mitigation-matrixfree}. 

Our complete results are visualized in Sec.~\ref{sec:mitigating-results}, here we describe the results in general.
When going from the raw to the mitigated results, we tracked the change in ground state probability, number of shots to reject fair sampling, and the slope of the line of best fit correlating the two.
We found mixed results in terms of ground state probability and fairness. 
Averaged over all problems, ground state probability improved by an average of $2.5\%$.
Meanwhile, the mean number of shots to reject fair sampling decreased by an average of $19.1\%$.
Finally, we observed that the slope of the line of best fit increased on 9 out of 10 problems, with an average improvement of $79.8\%$.
These results show that accounting for measurement error slightly improves ground state probability at the expense of fairness. 
The increase in slope suggests that the mitigated results are more dominated by structured error than the raw data.

This result may appear somewhat surprising, as measurement error on these devices is itself a structured error: qubits prepared in the 0 state are far less likely to be measured as 1 than vice versa. 
However, in the raw data we observe that ground states that are composed mostly of 0 qubits are disfavored by an average of $2\%$ (i.e. they appear $2\%$ less frequently than expected from a perfectly fair sampler). 
Meanwhile, states that are composed of a majority of 1 qubits are favored by an average of $3.4\%$.
It is unclear what errors are occuring during the evaluation of these circuits which lead to this bias in favor of 1, however it is likely measurement error is in fact helping results appear more fair by creating a slight counterbalancing bias in favor of 0. 
Indeed, after imposing our measurement error mitigation scheme we see that ground states that are composed mostly of 0 qubits are disfavored by an average of $5.9\%$, and those with mostly 1 qubits are favored by an average of $8\%$.
By accounting for the measurement bias in favor of 0, we have highlighted the fact that at least some of the purported fairness of these devices was in fact due to multiple biases partially cancelling each other out.  

We see from this simple example that fair results can be produced by three different quantum computers: those dominated by random noise, those with biased errors that nearly cancel each other out, and those with low error rates.
When determining the quality of these and future quantum computers, detailed analysis such as this is helpful in understanding which of the above scenarios best describes the device at hand. 

\section{Conclusion}
We have studied the extent to which today's gate-based quantum computing hardware can fairly sample from the space of optimal solutions for discrete optimization problems. 
First, we introduced a novel metric for evaluating the fairness of current hardware: the number of shots to reject fair sampling.
Using this metric, we found that hardware errors had a significant impact on fairness.
For short circuits with $\le 3$ qubits, higher fidelity qubits resulted in increased fairness. 
We call these circuits ``soft.''
However, for longer circuits with $\ge 4$ qubits, higher fidelity qubits actually resulted decreased fairness. 
We call these circuits ``hard.''
This is because low fidelity hardware begins behaving like a random number generator for long circuits, which contains no information but is relatively fair.

It would be interesting to test multiple quantum computing hardware technologies (e.g. optical, trapped ions, etc\ldots) to see if they have different fairness profiles.
Our studies show that individual machines on the IBM Q backend can have dramatically different fairness and ground state probabilities when different qubits are used.
As quantum computing hardware improves, we hope and expect to see an increase in qubit fairness and consistency.

In this study we have also briefly explored the ways in which mitigating errors can affect fairness.
Specifically, by we observed that measurement error was partially correcting other structured errors.
Measurement error mitigation therefore decreased fairness while modestly increasing ground state probability.
We believe that studying fairness in addition to ground state probability can give a more nuanced picture of the errors affecting NISQ hardware. 
This is important for two reasons.
First, random benchmark circuits are less susceptible to structured errors than circuits designed to solve realistic problems. 
And second, reported calibration and error rates reported are known to present an incomplete picture of how ordered circuits will perform~\cite{proctor2020measuring}.
This can be slightly improved upon by employing just-in-time noise data~\cite{wilson2020justintime}.
However, a broader theory of how different types of errors (read-out, crosstalk, etc.) affect not only ground state probability but also fairness is worth future study. 
This may involve utilizing different circuits or more refined statistical tools to highlight individual sources of error. 

\appendix
\section{IBM Q Backend Information}

\begin{table}[ht!]
\centering
\begin{tabular}{r|c|c|c|c|c|c|c|c|}
	\cline{2-9}
    & Backend & QV & Qubits & 2L & 3L & 4L & 4T & 5T  \\ 
    \cline{2-9} \noalign{\vskip\doublerulesep \vskip-\arrayrulewidth} \cline{2-9}
 	\multirow{10}{*}{October 2020} & 
    \verb|manhattan| & 32 & 65 & 72 & 190 & 232 & 96 & 92\\\cline{2-9}
    & \verb|montreal| & 32 & 27 & 28 & 74 & 80 & 48 & 36\\\cline{2-9}
    & \verb|toronto| & 32 & 27 & 28 & 74 & 80 & 48 & 36\\\cline{2-9}
    & \verb|rome| & 32 & 5 & 4 & 6 & 4 & 0 & 0\\\cline{2-9}
    & \verb|santiago| & 32 & 5 & 4 & 6 & 4 & 0 & 0\\\cline{2-9}
    & \verb|bogota| & 32 & 5 & 4 & 6 & 4 & 0 & 0\\\cline{2-9}
    & \verb|valencia| & 16 & 5 & 4 & 8 & 4 & 6 & 2\\\cline{2-9}
    & \verb|vigo| & 16 & 5 & 4 & 8 & 4 & 6 & 2\\\cline{2-9}
    & \verb|melbourne| & 8 & 15 & 20 & 72 & 130 & 66 & 120\\\cline{2-9}
    & \verb|ourense| & 8 & 5 & 4 & 8 & 4 & 6 & 2\\
    \cline{2-9} \noalign{\vskip\doublerulesep \vskip-\arrayrulewidth} \cline{2-9}
    \multirow{10}{*}{August 2021} 
	& \verb|montreal| & 128 & 27 & 28 & 74 & 80 & 48 & 36\\\cline{2-9}
	& \verb|mumbai| & 128 & 27 & 28 & 74 & 80 & 48 & 36\\\cline{2-9}
	& \verb|manhattan| & 32 & 65 & 72 & 190 & 232 & 96 & 92\\\cline{2-9}
	& \verb|brooklyn| & 32 & 65 & 72 & 190 & 232 & 96 & 92\\\cline{2-9}
	& \verb|toronto| & 32 & 27 & 28 & 74 & 80 & 48 & 36\\\cline{2-9}
	& \verb|sydney| & 32 & 27 & 28 & 74 & 80 & 48 & 36\\\cline{2-9}
	& \verb|guadalupe| & 32 & 16 & 16 & 40 & 40 & 24 & 16\\\cline{2-9}
    & \verb|bogota| & 32 & 5 & 4 & 6 & 4 & 0 & 0\\\cline{2-9}
	& \verb|lagos| & 32 & 7 & 6 & 14 & 8 & 12 & 4\\\cline{2-9}
	& \verb|jakarta| & 16 & 7 & 6 & 14 & 8 & 12 & 4\\\cline{2-9}
	& \verb|lima| & 8 & 5 & 4 & 8 & 4 & 6 & 2\\\cline{2-9}
\end{tabular}
	\caption{Details of IBM Q backends to be studied. 
	The name of the backend, the Quantum Volume (QV), the number of qubits, and the number of subsets of qubits on each device matching the relevant circuit architectures (2L, 3L, 4L, 4T, and 5T) are listed.}
	\label{tab:backends}
\end{table}
\clearpage
\section{Algorithm for Number of Shots to Reject Fair Sampling}\label{sec:NSRFS}
\begin{algorithm}[ht!]
    \caption{Number of Shots to Reject Fair Sampling} \label{alg:NSRFS}
    \begin{algorithmic}[1]
        \Require  Qubits $Q$, quantum circuit $C$ for a problem $P$ with $k$ degenerate ground states labeled $s_1,\ldots,s_k$; $n_s$ number of shots, $n_i$ number of inner loops
        \Ensure  Number $N$ of shots to reject fair sampling for $C$ on $Q$.
        \State  Execute $n_s$ shots of circuit $C$ on qubits $Q$ 
        \State  $o_i:=$ the number of times ground state $s_i$ observed
        \State  $w_i:= o_i/\sum_{i=1}^k o_i$ \Comment{relative frequency of ground state $s_i$}
        \State chiSqToReject := the $\chi^2$ value necessary to reject the fair sampling hypothesis with $95\%$ statistical significance for $k-1$ degrees of freedom
        \Function{medianChiSq}{$N$}
        	\State chiSqArray := \{\}
 			\While{\texttt{length}(chiSqArray) $< n_i$} 
				\State testSamples := $N$ samples drawn from the set $\{s_i\}$ with weights $\{w_i\}$
				\State testChiSq := $\chi^2$ for testSamples
				\State \verb|append|(chiSqArray, testChiSq)
			\EndWhile
			\State \textbf{return} \verb|median|(chiSqArray)
		\EndFunction
        \State  $N:= 2$ \Comment{$\chi^2$ ill-defined on sets of length 1}
        \While{$\Call{medianChiSq}{N} <$ chiSqToReject}
        \State  $N := 2N$
        \EndWhile
        \State upperBound := $N$
        \State lowerBound := $N/2$
        \While{upperBound-lowerBound $>2$}
        	\State $N := \left\lfloor(\text{upperBound}+\text{lowerBound})/2 \right\rfloor$
        	\If{$\Call{medianChiSq}{N} <$ chiSqToReject}
        	\State lowerBound := $N$
        	\Else
        	\State upperBound := $N$
        	\EndIf
        \EndWhile
        \State \textbf{return} $N$
    \end{algorithmic}
\end{algorithm}
For the calculations in this paper we used $n_s = 40960$ and $n_i=1000$.
As explained in the main text, these values were chosen via a `goldilocks' principle: much smaller, and the stochastic nature of the experiments resulted in inconsistent results, much larger and the computational burden was considerable. 
It is also worth noting that this algorithm becomes less useful as the samples become more evenly distributed, e.g. calculating the number of shots to reject fair sampling on a coin biased 50.001/49.999 takes $\mathcal{O}$(1hr).

\section{Details on Circuit Compilation}\label{sec:circ-details}
In this appendix we provide supplementary information detailing the specifics of our compilation techniques.
Full details of every hand compilation procedure would be too lengthy, instead we illustrate the compilation of Problem (d) as an example.
Before we delve into details, we mention some broad ideas which aided in the compilation.

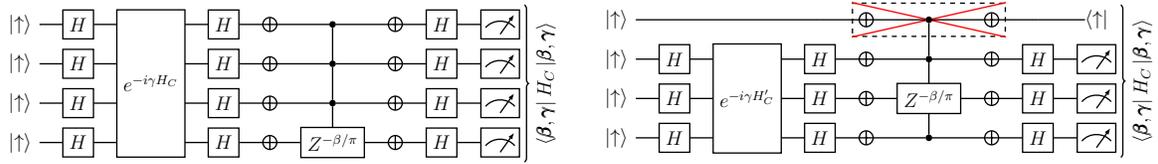
\begin{figure*}[ht!]
\centering
	\begin{adjustbox}{width=0.49\linewidth}
	\newcommand{\zgate}[1]{\gate{Z^{-\beta_{#1}/\pi}}}	
	\begin{quantikz}[row sep={24pt,between origins},execute at end picture={}]
		\lstick{\ket{\uparrow}}	
		& \gate{H}
		& \gate[4]{e^{-i\gamma H_C}}
		& \gate{H}
		& \targ{}
		& \ctrl{1}
		& \targ{}
		& \gate{H}
		& \meter{}\rstick[4]{\rotatebox{90}{$\bgstateT H_C \bgstate$}}
		\\	
		\lstick{\ket{\uparrow}}	& \gate{H}	&& \gate{H}	& \targ{} 	& \ctrl{1}	& \targ{} 	& \gate{H}	&\meter{}		\\	
		\lstick{\ket{\uparrow}}	& \gate{H}	&& \gate{H}	& \targ{}	& \ctrl{1}	& \targ{}	& \gate{H}	&\meter{}		\\	
		\lstick{\ket{\uparrow}}	& \gate{H}	&& \gate{H}	& \targ{}	& \zgate{}	& \targ{}	& \gate{H}	&\meter{}	
	\end{quantikz}
	\end{adjustbox}
	\hfill
	\begin{adjustbox}{width=0.49\linewidth}
	\newcommand{\zgate}[1]{\gate{Z^{-\beta_{#1}/\pi}}}	
	\begin{quantikz}[row sep={24pt,between origins},execute at end picture={}]
		\lstick{\ket{\uparrow}}	
		& \qw
		& \qw
		& \qw
		& \targ{}
		\gategroup[1,steps=3,style={dashed, inner sep=0pt},background]{}
		\gategroup[1,steps=3,style={cross out, red, inner sep=0pt},background]{}	
		& \ctrl{1}
		& \targ{}
		& \qw
		& \qw\bra{\uparrow}\rstick[4]{\rotatebox{90}{$\bgstateT H_C \bgstate$}}
		\\	
		\lstick{\ket{\uparrow}}	& \gate{H}	&\gate[3]{e^{-i\gamma H_C'}}
							& \gate{H}	& \targ{} 	& \ctrl{1}	& \targ{} 	& \gate{H}	&\meter{}	\\	
		\lstick{\ket{\uparrow}}	& \gate{H}	&& \gate{H}	& \targ{}	& \zgate{}	& \targ{}	& \gate{H}	&\meter{}	\\	
		\lstick{\ket{\uparrow}}	& \gate{H}	&& \gate{H}	& \targ{}	& \ctrl{-1}	& \targ{}	& \gate{H}	&\meter{}
	\end{quantikz}
	\end{adjustbox}
	\caption{(left) A 1-level Grover Mixer QAOA for an unconstrained optimization problem, 
		$U_S\ket{\uparrow^n} = \ket{\rightarrow^n}$, $U_M(\beta) = e^{-i\beta \ket{\rightarrow^n}\bra{\rightarrow^n}}$, and the Hamiltonian 
		$\ket{\rightarrow^n}\bra{\rightarrow^n} \cong \Id + \sum_i X_i + \sum_{\langle i,j \rangle} X_i X_j + \sum_{\langle i,j,k \rangle} X_i X_j X_k + \ldots$ 
		which has also been proposed to achieve fair sampling in quantum annealing~\cite{Matsuda2009}, 
		going beyond limited higher-order terms.\newline
		(right) For an unconstrained Ising problem with \emph{only quadratic terms} in the Hamiltonian 
		$H_C$ on $q_0, \ldots, q_{n-1}$, we break symmetries by setting $q_0 = \ket{\uparrow}$ and 
		$U_S\ket{\uparrow^n} = \Id \otimes H^{\otimes n-1} \ket{\uparrow^n} = \ket{\uparrow \rightarrow^{n-1}}$. 
		The quadratic terms of $H_C$ involving $q_0$ turn into linear terms of a Hamiltonian $H_C'$ on $q_1, \ldots, q_{n-1}$, 
		thereby reducing $q_0$ to a classical control bit. 
		The multi-control-$Z^{-\beta/\pi}$ gate is fully symmetric, thus we may swap controls and target.
	}
	\label{fig:qaoa}
\end{figure*}

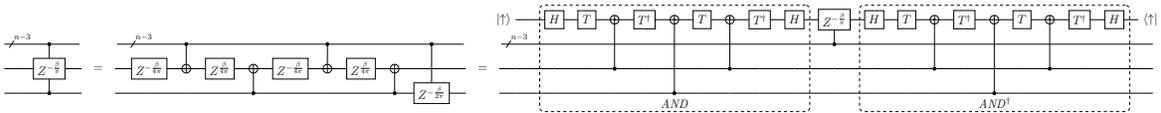
\begin{figure*}
	\centering
	\begin{adjustbox}{valign=T,width=\linewidth}
	\newcommand{\zgate}[2]{\gate{Z^{{#1}\frac{\beta}{{#2}\pi}}}}		
	\begin{quantikz}[row sep={24pt,between origins}, execute at end picture={
				\node[fit=(\tikzcdmatrixname-1-17)(\tikzcdmatrixname-4-25),draw,dashed,thick,rounded corners,inner xsep=10pt,inner ysep=12pt,xshift=5pt,yshift=-6pt,label={[yshift=16pt]below:$\mathit{AND}$}] {};
				\node[fit=(\tikzcdmatrixname-1-27)(\tikzcdmatrixname-4-35),draw,dashed,thick,rounded corners,inner xsep=10pt,inner ysep=12pt,xshift=5pt,yshift=-6pt,label={[yshift=16pt]below:$\mathit{AND}^{\smash{\dagger}}$}] {};
		}]
		&		&		&					&		&		&		&		&		&		&		&		&		&			& \lstick{\ket{\uparrow}}	& \qw	& \gate{H}	& \gate{T}	& \targ{}	& \gate{T^{\dagger}}	& \targ{}	& \gate{T}	& \targ{}	& \gate{T^{\dagger}}	& \gate{H}	& \zgate{-}{}	& \gate{H}	& \gate{T}	& \targ{}	& \gate{T^{\dagger}}	& \targ{}	& \gate{T}	& \targ{}	& \gate{T^{\dagger}}	& \gate{H}	& \qw \rstick{\bra{\uparrow}}		\\		
		& \qwbundle{n-3}& \ctrl{2}	& \qw\midstick[3,brackets=none]{=}	& \qwbundle{n-3}& \ctrl{1}	& \qw		& \qw		& \qw		& \ctrl{1}	& \qw		& \qw		& \ctrl{2}	& \qw\midstick[3,brackets=none]{=}	& \qwbundle{n-3}& \qw	& \qw		& \qw		& \qw		& \qw			& \qw		& \qw		& \qw		& \qw			& \qw		& \ctrl{-1}	& \qw		& \qw		& \qw		& \qw			& \qw		& \qw		& \qw		& \qw			& \qw		& \qw	& \qw		\\	
		& \qw		& \zgate{-}{}	& \qw					& \zgate{-}{4}	& \targ{}	& \zgate{}{4}	& \targ{}	& \zgate{-}{4}	& \targ{}	& \zgate{}{4}	& \targ{}	& \qw		& \qw					& \qw		& \qw	& \qw		& \qw		& \ctrl{-2}	& \qw			& \qw		& \qw		& \ctrl{-2}	& \qw			& \qw		& \qw		& \qw		& \qw		& \ctrl{-2}	& \qw			& \qw		& \qw		& \ctrl{-2}	& \qw			& \qw		& \qw	& \qw		\\	
		& \qw		& \ctrl{}	& \qw					& \qw		& \qw		& \qw		& \ctrl{-1}	& \qw		& \qw		& \qw		& \ctrl{-1}	& \zgate{-}{2}	& \qw					& \qw		& \qw	& \qw		& \qw		& \qw		& \qw			& \ctrl{-3}	& \qw		& \qw		& \qw			& \qw		& \qw		& \qw		& \qw		& \qw		& \qw			& \ctrl{-3}	& \qw		& \qw		& \qw			& \qw		& \qw	& \qw		
	\end{quantikz}
	\end{adjustbox}
	\caption{Two compilations of a large multi-control phase-shift gate (left) to gates with fewer controls:
		(middle) Direct decomposition into single-qubit phase-shift gates, large Toffolis, CNOTs, and a smaller multi-control phase-shift gate.
		(right) Decomposition using a $\ket{\uparrow}$-initialized ancilla into which we compute an AND (i.e., a Toffoli with mismatched phases),
		followed by a smaller multi-control phase-shift gate and an uncomputation of the AND.
	}
	\label{fig:multi-control-compilation}
\end{figure*}

\begin{itemize}
	\item Circuit parameters:
	The values for $\bbeta$ and $\bgamma$ in Table~\ref{tab:parameters} were found by 
	conducting a grid search over the angles $(\beta,\gamma) \in [-\pi,0) \times [-\pi,\pi)$ with a resolution of $\pi/60$
	and calculating the expectation value $\bgstateT H_C \bgstate$ classically 
	(where the search space for $\beta$ was cut in half due to a $(\beta,\gamma) \cong (-\beta,-\gamma)$ symmetry of the expectation value).
	This was efficient due to the relatively low circuit depths involved; for more complex circuits the hybrid approach detailed in Sec.~\ref{sec:review-QAOA} is more appropriate.

	\item Multi-controlled phase-shifts:
	First, we note that since terms in an Ising Hamiltonian pairwise commute, implementing the phase separating unitary can be done for each term individually, with 2 CNOTs and one phase-shift gate $Z^{\mp 2\gamma/\pi}$. 
	Care has to be taken, however, to address the connectivity constraints in the architectures. 
	More difficult is the compilation of the multi-control phase-shift gate at the heart of the Grover Mixer.
	Fig.~\ref{fig:multi-control-compilation} gives two recursive compilations into gates with fewer controls, one without ancillas, and one with a $\ket{\uparrow}$-initialized ancilla.
	We note that careful recursive calls to these compilations and using symmetries can result in some of the smaller multi-control gates to cancel each other. 

	\item SWAP operations: 
	We allow SWAP operations for qubits to change location in the circuit without having to go back to the original location. 
	This is possible because the Grover Mixer is fully symmetric, i.e. we can arbitrarily exchange controls with the phase-shift gate, thus the input order to the Grover Mixer does not matter. 
	We track qubit SWAPs with a permutation array from which in the end we read a read-out map for the measurements.
	This gives us a significant advantage over transpilation tools which are not aware of such symmetries.
	Introducing SWAPs (which usually cost 3 CNOTs) at the right place in a circuit (next to a CNOT) will cost only 1 CNOT. 
	Similarly, moving SWAPs through to the beginning or end of the circuit results in complete removal of the SWAP cost, as the SWAP gate can be replaced with a change in the permutation array and corresponding readout.
\end{itemize}

\subsection*{Example: Problem (d)}

Let us now describe in detail the compilation of Problem (d) to the 3L architecture.
Starting with the general form of the 1-level Grover Mixer QAOA with fixed qubit $q_0 := \, \uparrow$, we implement each 
quadratic term of the Ising Hamiltonian with 2 CNOTs and a single-qubit phase-shift gate 
$Z^{\mp 2\gamma/\pi} = \left(\begin{smallmatrix} 1&0\\0&e^{\mp i\gamma} \end{smallmatrix}\right)$.

\begin{figure*}
	\centering
	\begin{adjustbox}{width=\linewidth}
	\newcommand{\zgate}[1]{\gate{Z^{{#1}\frac{2\gamma}{\pi}}}}	
	\begin{quantikz}[row sep={24pt,between origins},execute at end picture={}]
		\lstick{$q_0\colon\ket{\uparrow}$}	
		& \qw
		& \gate[4][0pt]{{\rotatebox{90}{\large \quad$\substack{
			\mathclap{e^{-i\gamma (-Z_0 Z_1)}\cdot} \\[1ex]
			\mathclap{e^{-i\gamma (Z_1 Z_2 + Z_1 Z_3 + Z_2 Z_3)}}
		}$\quad}}}
		& \qw\midstick[4,brackets=none]{$\cong\ $}
		&
		& \lstick{$q_0\colon\ket{\uparrow}$}
		& \qw	& \qw	& \qw	& \qw	
		& \ctrl{1}	
		\gategroup[2,steps=1,style={dashed, inner sep=-5pt},background]{}
		\gategroup[2,steps=1,style={cross out, red, inner sep=-5pt},background]{}	
		& \qw	& \qw
		& \ctrl{1}
		\gategroup[2,steps=1,style={dashed, inner sep=-5pt},background]{}
		\gategroup[2,steps=1,style={cross out, red, inner sep=-5pt},background]{}
		& \qw	& \qw	& \qw	& \qw\rstick{$\bra{\uparrow}$}
		\\	
		\lstick{$q_1\colon\ket{\uparrow}$}	& \gate{H}	&&\qw	&& \lstick{$q_1\colon\ket{\uparrow}$}	& \gate{H}	& \ctrl{1}	& \qw		& \ctrl{1}	& \targ{}	& \zgate{-}	& \qw		& \targ{}	& \ctrl{1}	& \qw		& \ctrl{1}	& \qw\rstick{$q_1$}	\\	
		\lstick{$q_2\colon\ket{\uparrow}$}	& \gate{H}	&&\qw	&& \lstick{$q_2\colon\ket{\uparrow}$}	& \gate{H}	& \targ{}	& \zgate{}	& \targ{}	& \ctrl{1}	& \qw		& \targ{}
		\gategroup[2,steps=2,style={dashed,rounded corners,fill=blue!20, inner sep=0pt},label style={label position=below},background]{SWAP}
		& \ctrl{1}	& \targ{}	& \zgate{}	& \targ{}	& \qw\rstick{$q_3$}	\\	
		\lstick{$q_3\colon\ket{\uparrow}$}	& \gate{H}	&&\qw	&& \lstick{$q_3\colon\ket{\uparrow}$}	& \gate{H}	& \qw		& \qw		& \qw		& \targ{}	& \zgate{}	& \ctrl{-1}	& \targ{}	& \qw		& \qw		& \qw		& \qw\rstick{$q_2$}		
	\end{quantikz}
	\end{adjustbox}
	\caption{Phase Separator Compilation, Problem (d): Implementing the state preparation $U_S$ and the phase separator $U_P(\gamma) = e^{-i\gamma (-Z_0 Z_1 + Z_1 Z_2 + Z_1Z_3 + Z_2 Z_3)}$ 
		on a Linear Nearest Neighbor architecture. Quadratic terms commute pairwise, so they can be implemented individually with 2 CNOTs and one phase-shift gate each.
		Circuit size is reduced in three ways:
		(i) Setting $q_0 := \, \uparrow$ results in $q_0$ acting as a classical control bit, which can be removed.
		(ii) Directly combining a single quadratic-term phase separator with a SWAP gate increases the CNOT count by 1 (instead of the usual 3).
		(iii) We keep track of the resulting permutation of the qubits instead of undoing the SWAPs.
	}
	\label{fig:PS-compilation}
\end{figure*}
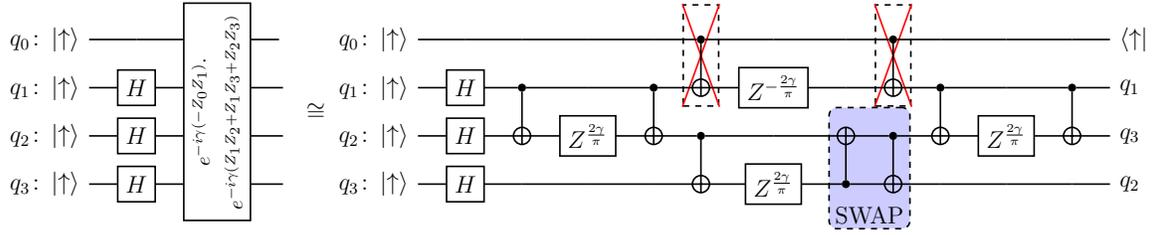

The CNOTs with control $q_0$ can be removed, leaving $q_0$ in the role of a classical bit, see Fig.~\ref{fig:PS-compilation}.
We notice that the other three quadratic terms form a triangle in the problem graph, hence we cannot map them to a 3L Linear Nearest Neighbor architecture, but rather need at least one SWAP. 
Inserting the SWAP as 3 CNOTs right to an existing CNOT of the phase separating unitary $e^{-i\gamma(Z_2 Z_3)}$ gives a cancellation of 2 CNOTs, thus the addition of the SWAP results in only 1 additional CNOT. 
Rather than reversing this SWAP at the end of $U_P(\gamma)$, we keep track of the permutation of the qubits ($q_2$ and $q_3$ have swapped places).

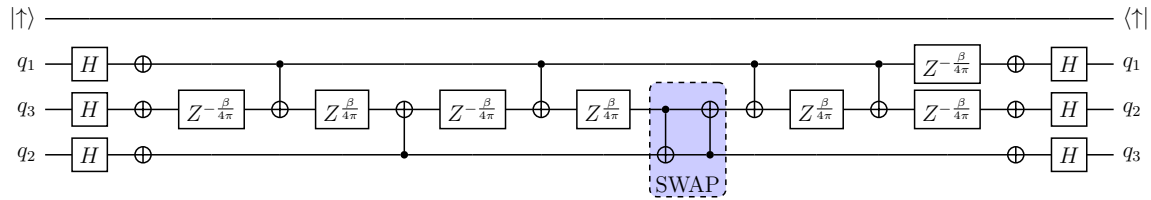
\begin{figure*}
	\centering
	\begin{adjustbox}{width=\linewidth}
	\newcommand{\zgate}[2]{\gate{Z^{{#1}\frac{\beta}{{#2}\pi}}}}	
	\begin{quantikz}[row sep={24pt,between origins},execute at end picture={}]
		\lstick{\ket{\uparrow}}	
		& \qw	& \qw	& \qw	& \qw	& \qw	& \qw	& \qw	& \qw	& \qw	& \qw	& \qw	& \qw	& \qw	& \qw	& \qw	& \qw	& \qw		
		& \qw\rstick{$\bra{\uparrow}$} \\	
		\lstick{$q_1$}	& \gate{H}	& \targ{}	&\qw		& \ctrl{1}	& \qw		& \qw		&\qw		& \ctrl{1}	& \qw		& \qw		& \qw		& \ctrl{1}	& \qw		& \ctrl{1}	& \zgate{-}{4}	& \targ{}	& \gate{H}	& \qw\rstick{$q_1$}	\\	
		\lstick{$q_3$}	& \gate{H}	& \targ{}	&\zgate{-}{4}	& \targ{}	& \zgate{}{4}	& \targ{}	&\zgate{-}{4}	& \targ{}	& \zgate{}{4}	& \ctrl{1}
		\gategroup[2,steps=2,style={dashed,rounded corners,fill=blue!20, inner xsep=0pt, inner ysep=5pt, yshift=-5pt},label style={label position=below},background]{SWAP}		
		& \targ{}	& \targ{}	& \zgate{}{4}	& \targ{}	& \zgate{-}{4}	& \targ{}	& \gate{H}	& \qw\rstick{$q_2$}	\\	
		\lstick{$q_2$}	& \gate{H}	& \targ{}	&\qw		& \qw		& \qw		& \ctrl{-1}	&\qw		& \qw		& \qw		& \targ{}	& \ctrl{-1}	& \qw		& \qw		& \qw		& \qw		& \targ{}	& \gate{H}	& \qw\rstick{$q_3$}		
	\end{quantikz}
	\end{adjustbox}
	\caption{Grover Mixer Compilation, Problem (d): Implementing the grover mixer unitary $U_M(\beta)$ 
		according to the high-level approch for multi-control phase-shift gates in Fig.~\ref{fig:multi-control-compilation}.
		Circuit size is reduced in two ways:
		(i) Moving SWAP gates through the circuit such that they appear at the beginning and the end, and thus can be tracked in the qubit permutation, 
		rather than implemented.
		(iii) Moving (multi-)control gates through the circuit such that they can be cancelled with adjoint gates (not applicable in this example). 
	}
	\label{fig:Mixer-compilation}
\end{figure*}

This end configuration (swapped qubits) now builds the start of the Grover Mixer unitary.
The $X$-gates and the control on qubit $q_0$ can be dropped, as explained in Fig.~\ref{fig:qaoa}~(right).
Using the recursive decomposition from Fig.~\ref{fig:multi-control-compilation}, we again see that we need to add a SWAP, see Fig.~\ref{fig:Mixer-compilation}.
The remaining single-controlled phase-shift gate between the middle two circuit wires can be implemented with 2 CNOTs and 3 phase-shift gates as shown.
Keeping track of the SWAP throughout the end of the circuit, we see that qubits $q_2$ and $q_3$ end up back in their original position -- in general, this need not be the case.

\subsubsection*{Comment on 5-qubit Ising Problems on 5T Using an Ancilla}

Since an ancilla offers no benefit for the implementation of the phase separating unitary, and due to limited connectivity in the considered architectures, we found that decomposing multi-controlled phase-shift gates using an ancilla as shown in Fig.~\ref{fig:multi-control-compilation}~(right) can only be used advantageously for the 5-qubit Ising Problems (a) and (b). 

Following the approach outlined for Problem (d), both Problems (a) and (b) result in a multi-controlled phase-shift gate acting on 4 qubits. 
Using a remaining qubit on the 5T architecture as a $\ket{\uparrow}$-initialized ancilla (which has to be swapped around), we can implement the mentioned decomposition, which should return the ancilla to the $\ket{\uparrow}$-state. 
As mentioned in the primary text, we discarded all samples where the ancilla qubit was not measured in the $\ket{\uparrow}$ state.
 \clearpage
\section{Measurement Error Mitigation Results}\label{sec:mitigating-results}

\begin{figure}[h!]
	\includegraphics[width=0.48\textwidth]{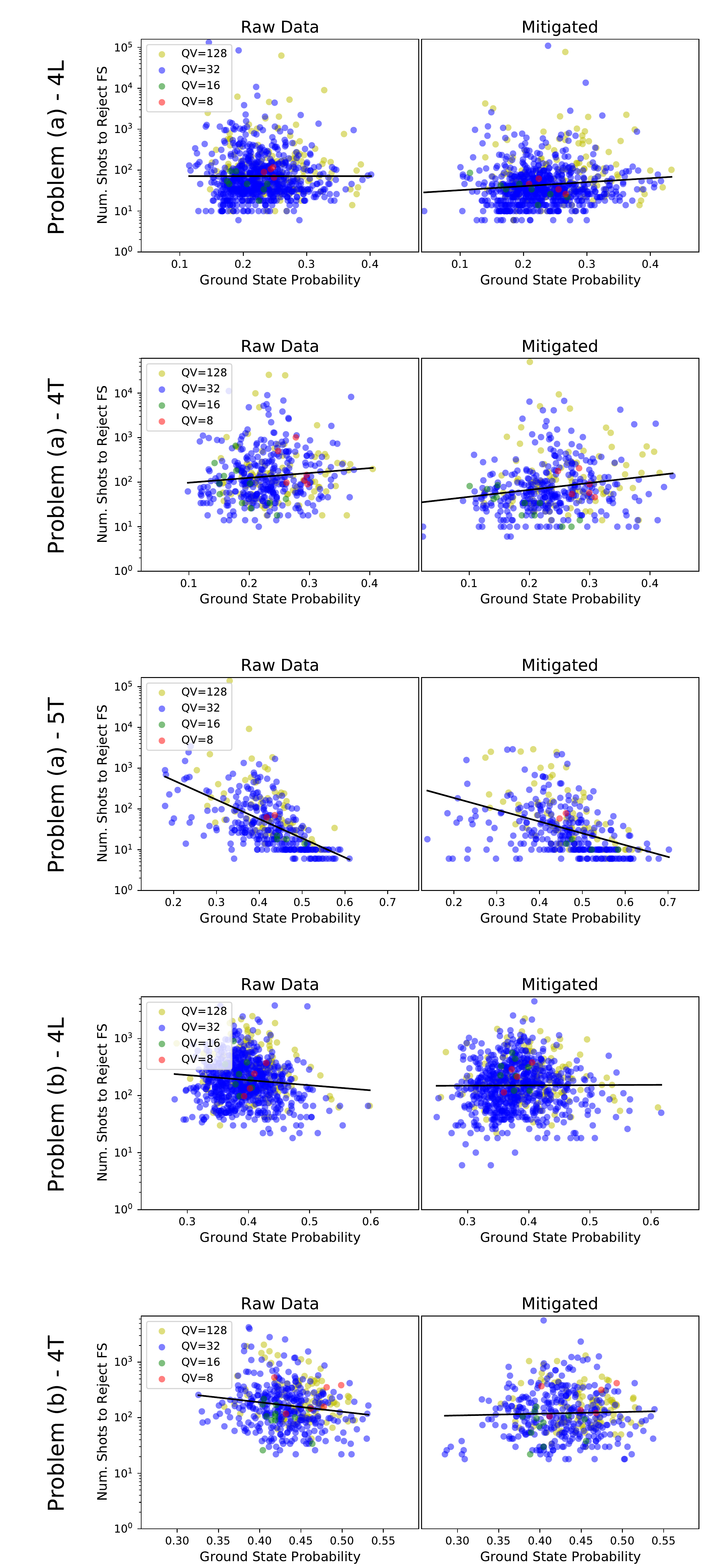}
	\includegraphics[width=0.48\textwidth]{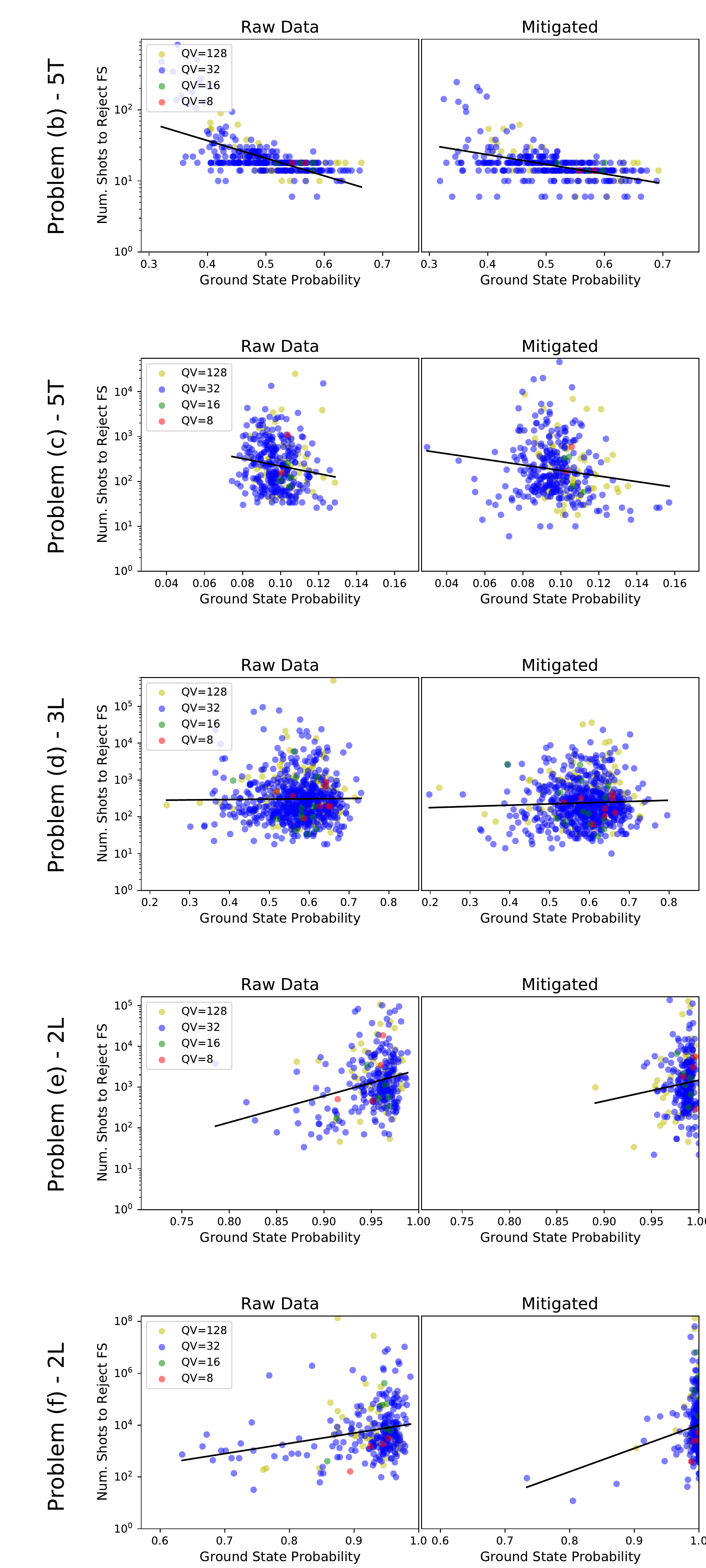}
	\caption{Results of measurement error mitigation across all problems. Ground state probability improved by an average of $2.5\%$, number of shots to reject fair sampling decreased by an average of $19.1\%$, and the slope of the line correlating the two increased by an average of $79.8\%$.}
\end{figure}

\clearpage
\bibliographystyle{plainurl}
\bibliography{manuscript}

\end{document}